\documentclass[superscriptaddress, amsmath,amssymb, aps, twocolumn,longbibliography]{revtex4-2}
\usepackage[plainpages = false, pdfpagelabels, 
                 bookmarks,
                 bookmarksopen = true,
                 bookmarksnumbered = true,
                 breaklinks = true,
                 linktocpage,
                 colorlinks = true,  
                 linkcolor = blue,
                 urlcolor  = blue,
                 citecolor = blue,
                 anchorcolor = green,
                 hyperindex = true,
                 hyperfigures
                 ]{hyperref} 
\usepackage{multirow}
\usepackage{enumerate}
\usepackage[caption=false,singlelinecheck=off]{subfig}
\usepackage{slashed}
\usepackage{braket}
\usepackage{subfloat}
\usepackage{bbold}
\usepackage{booktabs}
\usepackage{blindtext}
\usepackage{amsfonts}
\usepackage{bbm}
\usepackage{graphicx}
\usepackage{dcolumn}
\usepackage{bm}
\usepackage{amsmath, amssymb}
\usepackage[normalem]{ulem}
\usepackage{soul,xcolor}

\newcommand{\minus}{\scalebox{0.75}[1.0]{$-$}}
\setstcolor{red}

\begin{document}
 \title{Fingerprints of anti-Pfaffian topological order in quantum point contact transport}

\author{Jinhong Park}
\affiliation{\mbox{Institute for Quantum Materials and Technologies, Karlsruhe Institute of Technology, 76131 Karlsruhe, Germany}}
\affiliation{\mbox{Institut f{\"u}r Theorie der Kondensierten Materie, Karlsruhe Institute of Technology, 76131 Karlsruhe, Germany}}
\author{Christian Sp{\r a}nsl{\"a}tt}
\affiliation{\mbox{Department of Microtechnology and Nanoscience (MC2), Chalmers University of Technology, S-412 96 G\"oteborg, Sweden}}
\author{Alexander D. Mirlin}
\affiliation{\mbox{Institute for Quantum Materials and Technologies, Karlsruhe Institute of Technology, 76131 Karlsruhe, Germany}}
\affiliation{\mbox{Institut f{\"u}r Theorie der Kondensierten Materie, Karlsruhe Institute of Technology, 76131 Karlsruhe, Germany}}
\date{\today}
\begin{abstract}
    Despite recent experimental developments, the topological order of the fractional quantum Hall state at filling $\nu=5/2$ remains an outstanding question. We study conductance and shot noise in a quantum point contact device in the charge-equilibrated regime and show that, among Pfaffian, particle-hole Praffian, and anti-Pfaffian (aPf) candidate states, the hole-conjugate aPf state is unique in that it can produce a conductance plateau at $G=(7/3)e^2/h$ by two fundamentally distinct mechanisms. We demonstrate that these mechanisms can be distinguished by shot noise measurements on the plateaus. 
    We also determine distinct features of the conductance of the aPf state in the coherent regime.  Our results can be used to experimentally single out the aPf order. 
\end{abstract}
\maketitle

\textcolor{blue}{\noindent{\textit{Introduction.---}}}The fractional quantum Hall (FQH) state at filling $\nu=5/2$~\cite{Willet1987} is the prototypical candidate for a non-Abelian phase of matter~\cite{Moore1991}. This state has attracted immense attention as a tentative platform for topological quantum computations~\cite{Nayak2008}. However, to experimentally verify the realized topological order at this filling remains an outstanding problem in condensed-matter physics~\cite{Ma2019,Ma2022Aug}.

To describe the $\nu=5/2$ state, several candidate states were proposed, most prominently the Pfaffian (Pf)~\cite{Moore1991}, anti-Pfaffian (aPf)~\cite{Levin2007, Lee2007} and particle-hole Pfaffian (phPf)~\cite{Fidkowski2013, Zucker2016, Antonic2018} states, all with non-Abelian orders.  To date, numerical simulations favor either the aPf or Pf state~\cite{Wan2008, Rezayi2017, Rezayi2021}, while in GaAs/AlGaAs devices, recent measurements of the thermal conductance~\cite{Banerjee2018,Dutta2022Sep,paul2023topological} and upstream noise~\cite{Dutta2022} point towards the phPf state, supported by edge theory~\cite{Park2020OCt, Hein2023, Yutushui2023Dec}. 
Moreover, despite recent observations of several even-denominator states in novel 2D materials~\cite{Falson2015Apr,Zibrov2017Sep,Li2017Nov,Kim2019Feb,Shi2020Jul,Huang2022Jul,Hossain2023Mar,Chen2023Dec}, detailed transport experiments at $\nu=5/2$ in these materials remain elusive.

\begin{figure}[t!]
\includegraphics[width =1.0\columnwidth]{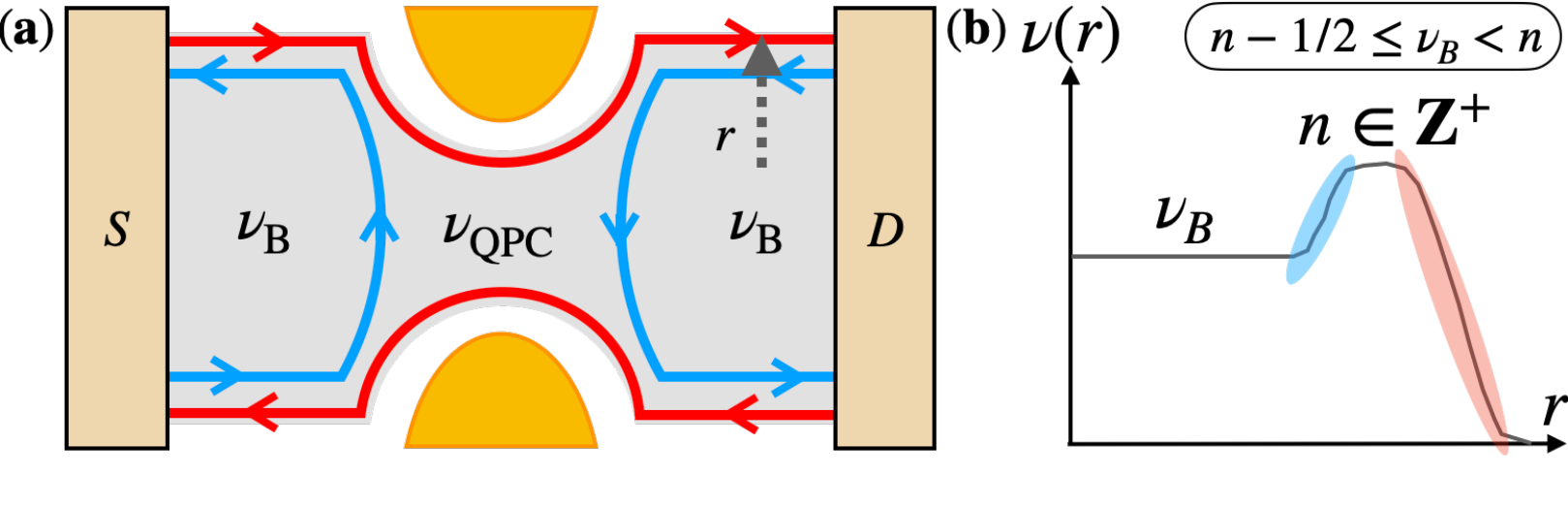}
\caption{{\bf (a) Schematic setup} to measure the two-terminal conductance $G\equiv I/(V_S-V_D)$ across a quantum point contact (QPC) in the FQH regime.  Here, $I$ is the current collected in drain ($D$), and $V_S$ and $V_D$ are the source ($S$) and drain voltages, respectively. For a hole-conjugate state, the QPC region can accommodate a FQH liquid with local filling $\nu_{\text{QPC}}$ \textit{higher} than the bulk filling factor $\nu_B$.  Red and blue solid lines with arrows depict counter-propagating edge modes.  {\bf (b) Sketch of the local filling factor} $\nu (r)$ along the gray dashed line in {\bf (a)}. Due to the hole-conjugate nature of the edge, there exists a region with higher, integer filling $n>\nu_B$. Red and blue jumps in $\nu(r)$ correspond to the edge modes in {\bf (a)}. }
\label{fig:IntroFig}
\end{figure}

In this paper, we address the $\nu=5/2$ conundrum by analyzing edge transport through a quantum point contact (QPC) device. Our main goal is to identify hallmarks of the aPf order related to its hole-conjugate nature, i.e., the presence of {\it counterpropagating} bosonic edge modes. In the regime of  equilibrated charge transport, the aPf state is expected to display a highly non-trivial plateau in the two-terminal conductance, $G=7/3$ (in units of $e^2/h$), when the QPC is continuously tuned by the corresponding gate voltage. This plateau arises when the local QPC  filling factor is $\nu_{\rm QPC}=3$
[Fig.~\ref{fig:IntroFig}\textcolor{blue}{(a)}], i.e., is \textit{higher} than the bulk filling $\nu_B=5/2$, see Ref.~\cite{Lai2013} for a discussion of related quantum-dot and line-junction setups. Among the non-Abelian candidates, this exotic mechanism of plateau formation is operative only for the aPf state due to its unique hole-conjugate character. However, 
a $G = 7/3$ plateau may form by another mechanism for \textit{any} $\nu = 5/2$ candidate state. This happens if the QPC instead \textit{lowers} the local density to the stable FQH filling $\nu_{\rm QPC}=7/3$, i.e., for $\nu_{\rm QPC}<\nu_B$ \cite{Spanslatt2020}.  We demonstrate that these two kinds of $G=7/3$ plateaus can be distinguished by electrical shot noise measurements, which thereby provide a unique fingerprint for the aPf state. Further, we explore the evolution of 
the $G=7/3$ plateau arising from  $\nu_{\rm QPC}=3$ in the aPf state in the regime of coherent charge transport, which can be reached for the lowest temperatures and bias voltages. We show that, due to disorder, $G$ then generically fluctuates with changing QPC gate voltage within the range $35/17\leq G\leq 3$. Among the three non-Abelian candidate states, $G > 5/2$ is reachable only for the aPf state, which thus provides a complementary fingerprint of this topological order in the QPC transport. 

\textcolor{blue}{\noindent{\textit{QPC conductance plateaus.---}}}The unusual situation with $\nu_{\rm QPC}>\nu_B$ is a feature common to all hole-conjugate states, i.e., for fillings satisfying $n - 1/2\leq\nu< n$, with $n \in \mathbb{Z}^+$. Indeed, all such states can be viewed as FQH liquids formed by condensation of hole-like quasiparticles on top of an integer number $n$ of filled Landau levels~\cite{Haldane1983,Halperin1984a,MacDonald1990}. As a consequence, hole-conjugate states naturally accommodate local regions with $\nu (r) = n > \nu_B$ [see Fig.~\ref{fig:IntroFig}\textcolor{blue}{(b)}]. This property suggests the possibility of a local region with integer-valued filling $\nu_{\text{QPC}} = n$, despite the application of a negative gate voltage that normally \textit{lowers} the local density. Having $\nu_{\rm QPC}>\nu_B$, together with the assumption of fully equilibrated charge transport, leads to non-trivial conductance plateaus~\cite{SuppMat}
\begin{align} \label{eq:nontrivialconductanceplateau}
    G = \frac{\nu_B \nu_{\text{QPC}} - (2 \nu_B - \nu_{\text{QPC}}) \nu_T}{2 \nu_{\text{QPC}} - \nu_B - \nu_T}, \,\,\, \text{for}\,\,\, \nu_{\text{QPC}} > \nu_B\,,
\end{align}
see Fig.~\ref{fig:IntroFig}\textcolor{blue}{(a)} for the schematic QPC setup. 
Here, $\nu_T$ is the total filling factor discontinuity associated with fully transmitted modes (without any coupling to the other modes). Equation~\eqref{eq:nontrivialconductanceplateau} explains recently observed, unusual conductance plateaus for hole-conjugate FQH states~\cite{Nakamura2023Feb, Fauzi2023Jan, Yan2023Mar}. In particular, for the $\nu_B = 2/3$ state (with $\nu_{\text{QPC}} = 1$ and $\nu_T = 0$), a $G = 1/2$ plateau was recently observed~\cite{Nakamura2023Feb, Fauzi2023Jan}. Also FQH states in higher Landau levels were observed to display non-trivial plateaus classified by Eq.~\eqref{eq:nontrivialconductanceplateau}~\cite{Yan2023Mar}. 

Crucially, among the $\nu = 5/2$ candidate states, it is only the aPf state that is hole-conjugate. Hence, only the aPf can produce a conductance plateau by the mechanism governing Eq.~\eqref{eq:nontrivialconductanceplateau}. For $\nu_B=5/2$, $\nu_T = 2$, and $\nu_{\text{QPC}}=3$, Eq.~\eqref{eq:nontrivialconductanceplateau} evaluates to $G = 7/3$, in agreement with Ref.~\cite{Lai2013}.

However, conductance plateaus may also arise for a reduced density in the QPC region~\cite{Spanslatt2020}, with
\begin{align} \label{eq:trivialconductanceplateau}
    G = \nu_{\text{QPC}}, \quad \text{for}\,\,\, \nu_{\text{QPC}} < \nu_B\,.
\end{align}
In contrast to the  plateaus~\eqref{eq:nontrivialconductanceplateau}, Eq.~\eqref{eq:trivialconductanceplateau} holds for any FQH state provided that the state with filling $\nu_{\text{QPC}} < \nu_B $ is stabilized in the QPC region. Experimental observations of plateaus for various FQH states~\cite{Bid2009shot,Bhattacharyya2019,Biswas2022shot} can be attributed to this mechanism. 
We see that, according to Eq.~\eqref{eq:trivialconductanceplateau}, the value $G=7/3$ is also generated for $\nu_B=5/2$ regardless of the bulk topological order if the QPC region hosts a $\nu_{\text{QPC}} = 7/3$ FQH state. Such a state is indeed the most prominent and stable state in the range $2< \nu < 5/2$. Hence, to differentiate the two distinct types of 7/3 plateaus and thus to find unique fingerprints for the aPf state, complementary measurements are needed.  We will show that on-plateau shot noise measurements meet this demand. 


\begin{figure}[t!]
\includegraphics[width =0.9\columnwidth]{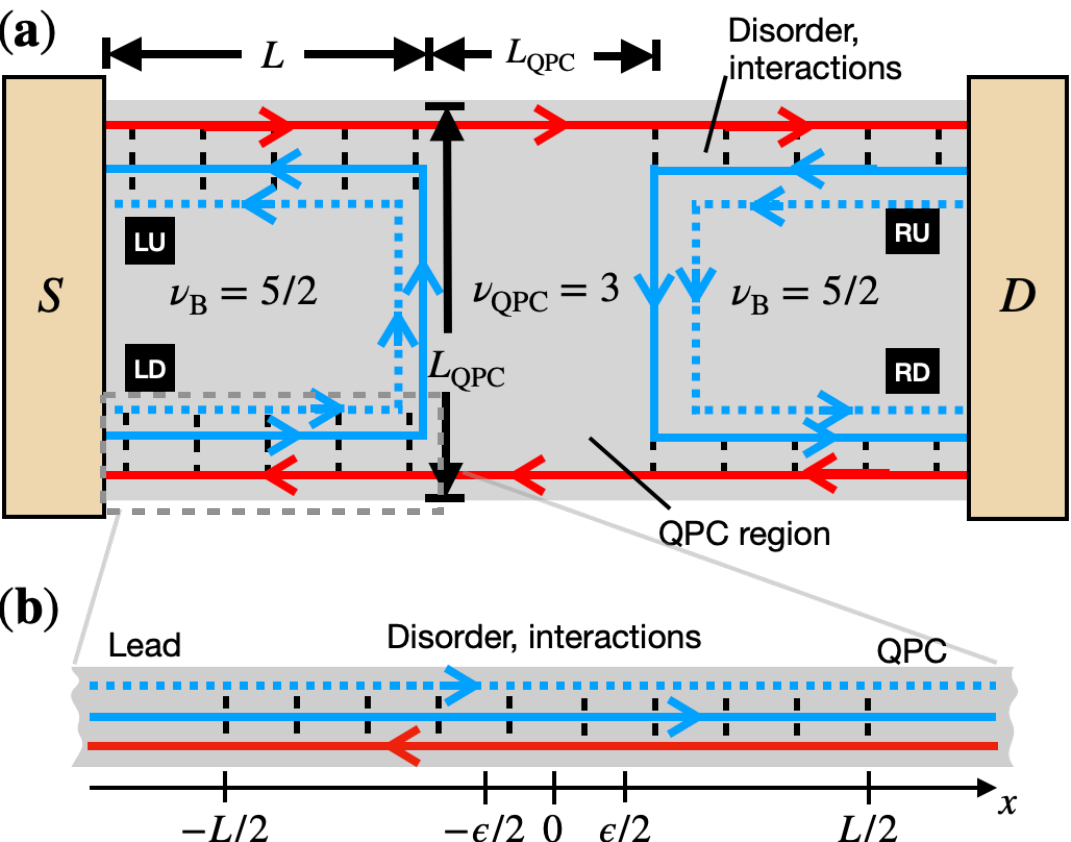}
\caption{{\bf (a) QPC Configuration} for the anti-Pfaffian state with QPC filling stabilized at $\nu_{\text{QPC}} = 3$. Red and blue solid lines depict $\delta \nu = 1$ and $\delta \nu = -1/2$ edge modes propagating in the opposite directions. The dashed blue line depicts a Majorana mode propagating in the same direction as the $\delta \nu = -1/2$ mode. The edge modes are coupled by disorder-induced scattering in each individual arm of the QPC. The two fully transmitted integer modes in the lowest Landau levels are not shown.
  {\bf (b) Blow-up of one QPC arm} which bridges the interaction and disorder-free lead region ($x < -L/2$) and the QPC region ($x > L/2$). All impact of disorder is accounted for by the narrow region $|x|<\epsilon/2$ with $\epsilon \rightarrow 0^+$. This feature holds for all four arms LU, LD, RU, and RD in panel {\bf (a)}.}
\label{fig:APF_Setup}
\end{figure}

\textcolor{blue}{\noindent{\textit{APf edge theory.---}}}The edge consists of three modes in the second Landau level: two counter-propagating bosonic modes $\phi_1$ 
(red solid lines in Fig.~\ref{fig:APF_Setup}) and $\phi_{\frac{1}{2}}$ (blue solid lines) associated with the filling factor discontinuities 
$\delta \nu = 1$ and $\delta \nu = -1/2$, respectively,
as well as one charge-neutral Majorana mode $\chi=\chi^\dagger$  (blue dashed lines)~\cite{Levin2007, Lee2007}. We disregard two integer modes of the lowest Landau levels, assuming that they are decoupled and simply give a contribution 2 to $G$. The edge action is $S= \int dt (\mathcal{L}_0 + \mathcal{L}_{\text{dis}} )$ with
\begin{align} \label{eq:bareaction}
   &\mathcal{L}_0 =  \int \frac{dx}{4\pi} \big [ \big (-\partial_x \phi_1 (\partial_t + v_1 \partial_x ) \phi_1 + 2\partial_x \phi_{\frac{1}{2}} (\partial_t - v_{\frac{1}{2}} \partial_x ) \phi_{\frac{1}{2}} \nonumber \\ & \quad \quad \quad - 2 v_{\text{int}} \partial_x \phi_1 \partial_x \phi_{\frac{1}{2}} \big)   - i \chi(\partial_t - v_{M} \partial_x ) \chi \big ] \,, \nonumber \\
    &\mathcal{L}_{\text{dis}} = - \frac{1}{\sqrt{2\pi a}}\int dx  \chi(x) \big ( \xi (x) e^{i (\phi_1 + 2 \phi_{\frac{1}{2}})} - h.c.  \big )\,.
\end{align}
 Here, $v_1$, $v_{\frac{1}{2}}$, and $v_M$ are the mode speeds, $v_{\rm int}$ is the inter-mode interaction strength, and $a$ is an ultraviolet length cutoff. The term $\mathcal{L}_{\text{dis}}$ describes disorder-induced electron tunneling with the random complex amplitude $\xi(x)$. The scaling dimension of the disorder term evaluated with respect to $\mathcal{L}_0$ is $\Delta =1/2+(3/2-2x)/\sqrt{1-2x^2}$, where $x = v_{\text{int}} / (v_1 + v_{\frac{1}{2}})$.
 
 When $\mathcal{L}_{\text{dis}}$ is relevant in the renormalization group (RG) sense, i.e., for $\Delta < 3/2$~\cite{Giamarchi1988}, the edge is driven to a disordered fixed point where $\Delta = 1$~\cite{Levin2007,Lee2007}. At this point, the edge hosts three decoupled modes: one charge mode $\phi_c\equiv \sqrt{2} (\phi_1 + \phi_{\frac{1}{2}})$, one neutral mode $\phi_n \equiv \phi_1 + 2\phi_{\frac{1}{2}}$, and the remaining Majorana mode $\chi$. These are described by $\mathcal{L}_{\text{fix}} = \mathcal{L}_c + \mathcal{L}_n$, where
 \begin{align} \label{eq:fixedpointLagrangian}
    &\mathcal{L}_c = \int \frac{ dx}{4\pi}  [-\partial_x \phi_c (\partial_t + v_c \partial_x ) \phi_c ] \,, \nonumber \\ 
    &\mathcal{L}_n =  \int  dx  \Big [\frac{1}{4\pi}\partial_x \phi_n (\partial_t - \bar{v}_n \partial_x ) \phi_n - i \chi (  \partial_t - \bar{v}_n \partial_x ) \chi \nonumber \\
    & \quad  \quad - \frac{1}{\sqrt{2 \pi a}}   \left ( \xi (x) e^{i \phi_n } - h.c.  \right) \chi \, \Big ]\,. 
\end{align}
Near the fixed point, all possible terms (e.g.,  $\phi_c\minus\phi_n$ interactions and the velocity anisotropy) are RG irrelevant~\cite{Levin2007,Lee2007}. Still, these terms cause decoherence which leads to inter-mode equilibration governed by the characteristic charge equilibration length $\ell_{\text{eq}}^{c}\sim T^{-2}$, with $T$ the temperature. This length defines two distinct charge transport regimes: the {\it coherent} regime $L \ll \ell_{\text{eq}}^{c}$ and the {\it incoherent} regime $L \gg \ell_{\text{eq}}^{c}$. $L$ denotes the length of the arms of the QPC (see Fig.~\ref{fig:APF_Setup}). In the following, we separately discuss transport properties in these two regimes. 

\begin{figure}[t!]
\includegraphics[width =0.99\columnwidth]{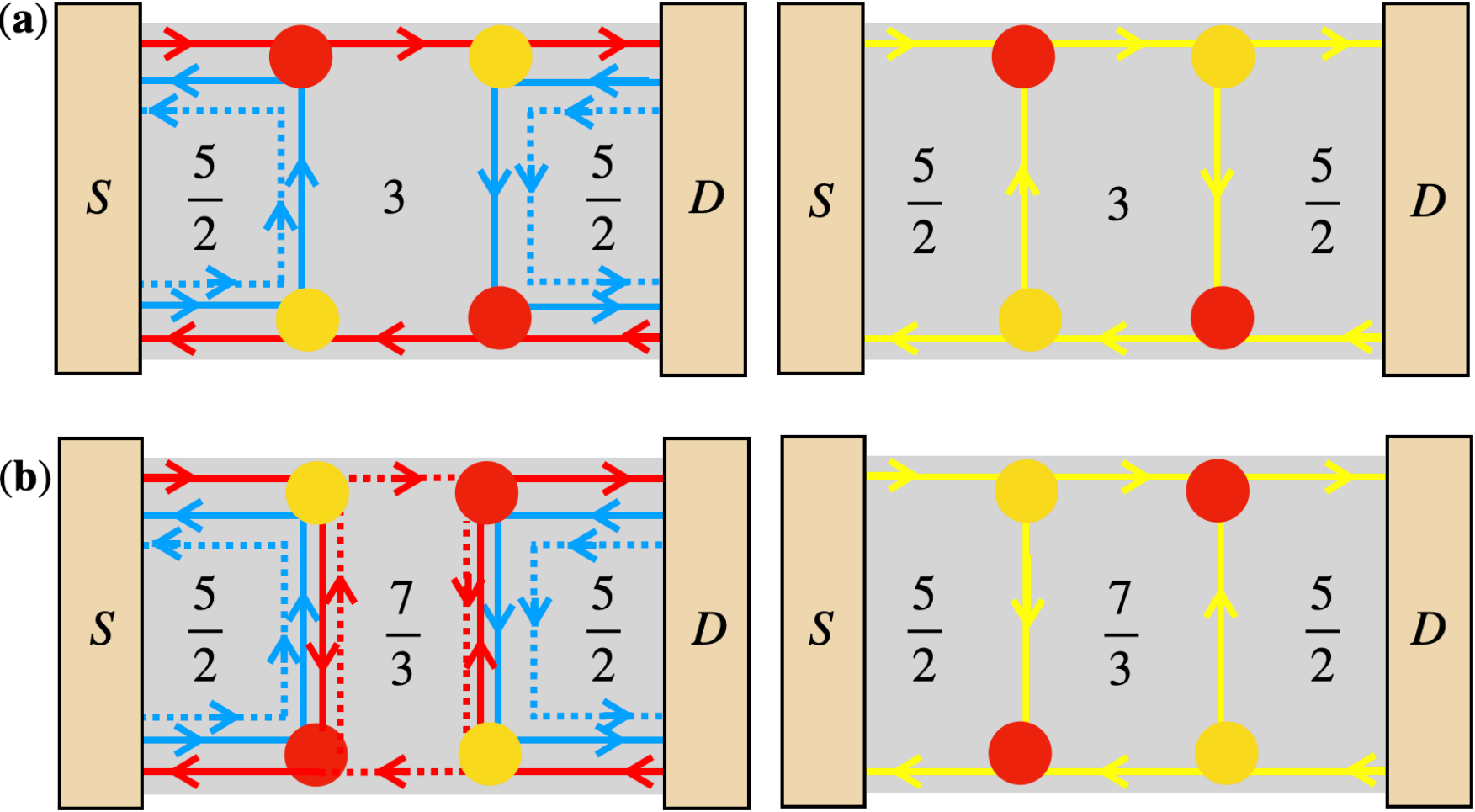}
\caption{{\bf  Hot spots (red regions) and noise spots (yellow regions)} for the two distinct $G = 7/3$ plateaus of the aPf state in the incoherent regime: {\bf (a)} $\nu_{\rm QPC} = 3 >\nu_B = 5/2$ and {\bf (b)} $\nu_{\rm QPC} = 7/3 < \nu_B$. Left panels describe the edge configurations and the right panels describe the charge flow (yellow, solid lines) along each edge segment.}
\label{fig:Noises}
\end{figure}

\textcolor{blue}{\noindent{\textit{Shot noise in the incoherent regime.---}}}We now show that the two types of $G=7/3$ plateaus for the aPf generate very distinct (dc) shot noise characteristics as they correspond to two distinct QPC configurations: (i) $\nu_{\rm{QPC}} = 3 > \nu_{B}=5/2$ in Fig.~\ref{fig:Noises}\textcolor{blue}{(a)} and (ii) $\nu_{\rm{QPC}} = 7/3 < \nu_{B}$ in Fig.~\ref{fig:Noises}\textcolor{blue}{(b)}. Measurements of non-equilibrium noise involving (partial) equilibration on edge segments were recently proposed and experimentally implemented as a versatile tool to probe the topological order in various FQH setups, including transport on single edge segments~\cite{Park2019,Spanslatt2019,Park2020OCt,Melcer2022Jan,Kumar2022Jan}, engineered interfaces~\cite{Dutta2022Sep,Hein2023, manna2023classification,Yutushui2023Dec}, and QPCs~\cite{Spanslatt2020,manna2023shot}.

The shot noise generated in the process of equilibration (a hallmark of the incoherent regime) among counter-propagating edge modes depends on the competition of several characteristic length scales. We therefore start by establishing the relevant hierarchy of these length scales. Motivated by recent experimental observations in both GaAs/AlGaAs and graphene devices, we assume 
\begin{align} 
\label{eq:lengthscales}
    \ell_{\text{eq}}^c \ll L_{\text{QPC}} \ll \ell_{\text{eq}}^h \ll L.
\end{align}
Here, $\ell_{\text{eq}}^h$ is the heat equilibration length
and $L_{\text{QPC}}$ is the size of the QPC region [see Fig.~\ref{fig:APF_Setup}\textcolor{blue}{(a)}]. While full charge equilibration is achieved over a very short scale ($\lesssim$1$\mu$m) in essentially all FQH devices 
(see, however, Refs.~\cite{Lafont2019, Cohen2019, Hashisaka2021,Hashisaka2023} for prominent exceptions), heat equilibration is often poor at low temperatures~\cite{Simon2020, Ma2020Jul, Asasi2020, Srivastav2021May,Breton2022, Kumar2022Jan, Srivastav2022Sep,Melcer2022Jan, Dutta2022Sep}. We also assume no edge reconstruction.

The mechanism for shot noise generation on a conductance plateau in the incoherent regime is due to charge partitioning at so-called ``noise spots'' (denoted by yellow regions in Fig.~\ref{fig:Noises})~\cite{Park2019, Spanslatt2019, Spanslatt2020,Park2020OCt}. An inter-mode charge tunneling event contributes to the noise only if the constituents of the resulting particle-hole pair reach different contacts $S$ and $D$. This indeed happens only at the noise spots, where the charge current partitions into these contacts, see the right panels in Fig.~\ref{fig:Noises} for depictions of the charge flows (solid, yellow lines) in the device.  Importantly, the tunneling processes in each noise spot are dominantly generated by an increase of the noise spot temperature. This heating occurs by heat flowing from hot spots (red regions in Fig.~\ref{fig:Noises}), where all Joule heating in the device occurs, to the noise spots. The generated noise in a noise spot, $S_{\rm NS}$, is given by
\begin{align} \label{eq:noiselinejunction}
    S_{\rm NS} = \frac{e^2}{h} \frac{(\nu_d - \nu_u) \nu_u}{\nu_d} k_B (T_{d}+ T_{u})\,.
\end{align}
Here, $\nu_{d/u} = \sum_{n=1}^{n_{d/u}} \delta \nu_{d/u, n}$ is the total filling factor discontinuity of the charged modes propagating downstream and upstream, respectively, and $n_d$ and $ n_u$ denote the number of charged downstream and upstream modes, respectively. Furthermore, $T_{d/u}$ is the temperature of all downstream and upstream modes at the noise spot. 

We next compute the total noise in the drain contact $D$. To this end, we calculate the local temperatures at the noise spots by formulating and solving a set of transport equations for the charge and energy currents along each edge segment in the device~\cite{SuppMat}. We supplement these equations by boundary conditions for the two contact voltages $V_S = V$, $V_D = 0$ and temperatures $T_S=T_D =0$. The boundary conditions connecting different edge segments follow from charge and energy conservation.

\begin{figure}[t!]
\includegraphics[width =0.99\columnwidth]{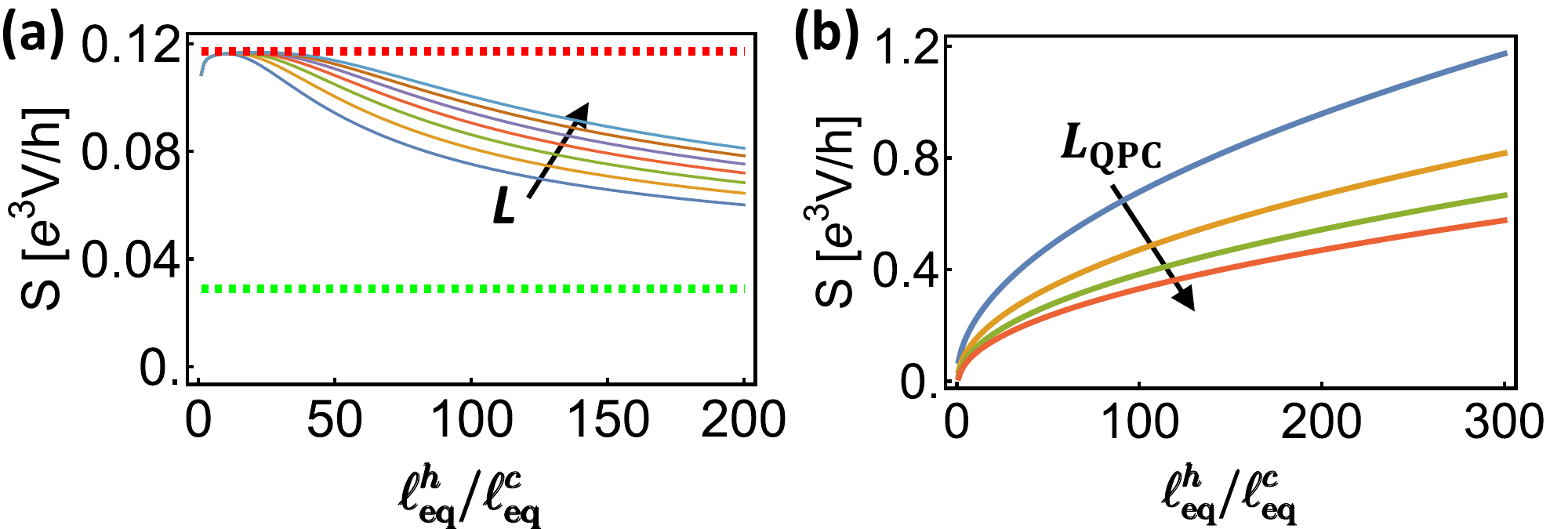}
\caption{{\bf  Anti-Pfaffian noise characteristics $S$} (in units of $e^3V/h$ where $V$ is the bias voltage) vs the ratio of heat and charge equilibration lengths $\ell_{\text{eq}}^h / \ell_{\text{eq}}^c$. The noise is evaluated for the  two different QPC configurations in Fig.~\ref{fig:Noises}. {\bf (a)} $S$ is plotted for $\nu_{\text{QPC}} = 3 > \nu_B = 5/2$. The solid lines correspond to different choices of QPC arm lengths $200\leq L/ \ell_{\text{eq}}^c\leq 500$ in steps of $50$. The red and green dashed lines correspond to analytical values obtained for the hierarchy \eqref{eq:lengthscales}, and in the limit $\ell_{\text{eq}}^c \ll L \ll \ell_{\text{eq}}^h$, respectively. {\bf (b)} $S$ is computed for $\nu_{\text{QPC}} = 7/3 < \nu_B $ for different $L_{\text{QPC}}$, $3 \leq L_{\text{QPC}} /\ell_{\text{eq}}^c  \leq 12$ in steps of $3$, with fixed $L = 300 \ell_{\text{eq}}^c$. 
}
\label{Fig:noisechar}
\end{figure}

In configuration (i), for the experimentally relevant length hierarchy \eqref{eq:lengthscales}, we obtain $k_B T_{d}= \sqrt{5/6}\,eV/ \pi$ and $k_B T_{u} = \sqrt{5}\, eV/ (3\pi) $. Then, the noise $S$ measured in the drain $D$ is given by 
\begin{align} \label{eq:noiseanalytic}
S = \frac{8S_{\rm NS}}{9}  = \frac{8e^3 (\nu_d - \nu_u) \nu_u}{9 h \nu_d} k_B (T_d+ T_u) \approx
\frac{0.12e^3V}{h}\,,
\end{align}
upon substituting $\nu_d = 1$, $\nu_u = 1/2$ for the aPf state. The prefactor $8/9$ stems from the conversion of the single noise spot $S_{\rm NS}$ to the total noise $S$~\cite{SuppMat}. 
In Fig.~\ref{Fig:noisechar}\textcolor{blue}{(a)}, we plot $S$ vs the ratio $\ell_{\text{eq}}^h/\ell_{\text{eq}}^c$ for different choices of $L$. For relatively small $\ell_{\text{eq}}^h / \ell_{\text{eq}}^c$, $S$ approaches Eq.~\eqref{eq:noiseanalytic}, as indicated by the red, dashed line. Further increasing $\ell_{\text{eq}}^h / \ell_{\text{eq}}^c$ causes $S$ to monotonously decrease towards $S \approx 0.029 e^3 V/ h$ (green, dashed line), i.e, the value in the regime of vanishing thermal equilibration in the QPC arms, $\ell_{\text{eq}}^h \gg L$~\cite{SuppMat}.  As $L$ increases, the region of Eq.~\eqref{eq:noiseanalytic} broadens, which reflects that $\ell_{\text{eq}}^h$ satisfies the hierarchy \eqref{eq:lengthscales} for a broader range of values.   

In configuration (ii), we find instead $S \propto \sqrt{\ell_{\text{eq}}^h / L_{\text{QPC}}}$ for the hierarchy~\eqref{eq:lengthscales}, so that $S$ increases as a function of $\ell_{\text{eq}}^h$, in contrast to
Eq.~\eqref{eq:noiseanalytic} [see Fig.~\ref{Fig:noisechar}\textcolor{blue}{(b)}]. 
This behavior results from an emergent, circulating $\delta \nu = 1/3$ mode in the QPC region [see Fig.~\ref{fig:Noises}\textcolor{blue}{(b)}]. This mode heats up continuously by Joule heating at the hot spots until the heat escapes to other edge modes by thermal equilibration. The circulating heat current effectively winds $\sim \ell_{\text{eq}}^h / L_{\text{QPC}}$ times before it reaches a steady state, and as a result, $T_d$, $T_u$, and $S$ scale as $\sqrt{\ell_{\text{eq}}^h / L_{\text{QPC}}}$. Crucially, regardless of the topological order of the $\nu = 5/2$ states, the $\delta \nu = 1/3$ mode always exists in configuration (ii), and therefore this type of noise characteristics holds for all $\nu = 5/2$ candidate states (including Abelian states). 

\textcolor{blue}{\noindent{\textit{Conductance in the coherent regime.---}}}At sufficiently low temperatures and voltages, FQH experiments can reach the regime where the charge equilibration is weak~\cite{Cohen2019,Hashisaka2021,Hashisaka2023}. In this regime for the aPf state, each QPC arm segment is described by the fixed point Lagrangian \eqref{eq:fixedpointLagrangian}. The neutral sector $\mathcal{L}_n$ has an emergent SO$(3)$ symmetry~\cite{Levin2007, Lee2007}: by defining the Majorana mode triplet $\psi^T \equiv (\psi_1, \psi_2, \psi_3)$ with 
$\psi_1 = \frac{1}{2 \sqrt{\pi a}} (e^{i \phi_n} + e^{-i \phi_n})$, $\psi_2 = \frac{-i}{2 \sqrt{\pi a}} (e^{i \phi_n} - e^{-i \phi_n})$ and $\psi_3 = \chi$, we express
$\mathcal{L}_n$ as
\begin{align} \label{eq:armso3}
   \mathcal{L}_n =  \int_{x \in \mathcal{R}_{\text{Arm}}} dx &\big [\minus i\psi^T (\partial_t - \bar{v}_n \partial_x + \sum_{a= 1,2} \xi_a \hat{L}_a ) \psi \big ]\,.
\end{align}
Here, $\xi_1 \equiv \text{Re}(\xi)$, $ \xi_2 \equiv - \text{Im}(\xi)$ and $(\hat{L}_a)_{bc} \equiv \epsilon_{abc}$ are the generators of SO$(3)$. The integration range in Eq.~\eqref{eq:armso3} includes all four arm regions in the QPC, $x \in \mathcal{R}_{\text{Arm}}$; i.e., the regions $|x| < L/2 $ in Fig.~\ref{fig:APF_Setup}\textcolor{blue}{(b)}. In contrast, the disorder-free lead and QPC regions are described by the clean $\mathcal{L}_0$ in Eq.~\eqref{eq:bareaction}. We next perform the transformation $\tilde{\psi} (x) = U^T (x) \psi (x)$ with the rotation matrix
\begin{align} \label{eq:orthogonalmatrix}
    U(x) = U_0(x, -\infty) \equiv T_x \exp \Big(\sum_{a=1}^2 \int_{-\infty}^x dx' \frac{\xi_a (x') \hat{L}_a}{\bar{v}_n} \Big)\,,
\end{align}
where $T_x$ denotes path-ordering in $x$. The disorder term for $x \in \mathcal{R}_{\text{Arm}}$ is then fully absorbed which makes the SO$(3)$ symmetry manifest~\cite{Levin2007}: 
\begin{align} \label{eq:armso3rotatedbasis}
   \mathcal{L}_n =  \int_{x \in \mathcal{R}_{\text{Arm}}} dx &\big [- i \tilde{\psi}^T (\partial_t - \bar{v}_n \partial_x) \tilde{\psi} \big ]\,.
\end{align}
In the presence of lead and QPC regions, the  transformation \eqref{eq:orthogonalmatrix} is inconvenient, as first pointed out in Ref.~\cite{Protopopov2017} for the $\nu = 2/3$ edge. To clarify why the same is true here, we first introduce an Euler-angle decomposition of the phase factor, $U_{\text{tot}}$, accumulated in the disordered region:
\begin{align} \label{eq:totalphasefactor}
U_{\text{tot}}  \equiv U_0(L/2, - L/2 )\equiv S_+ e^{\hat{\omega}} S_{-}^{-1}\,.
\end{align}
Here, $S_+$ and $S_-$ denote two different rotations around the $\hat{z}$ axis (in the Majorana triplet space), while $e^{\hat{\omega}}$ rotates around the $\hat{x}$ axis by the matrix $\hat{\omega} = \theta \hat{L}_1$ with angle $0 \leq \theta \leq \pi$. Then, $e^{\hat{\omega}}$ and also $U_\text{tot}$ rotates $\psi(x)$ around the $\hat{x}$ axis in both lead and QPC regions (i.e., for $|x|>L/2$). These rotations produce non-trivial vertex operators $e^{\pm i \phi_n}$ in these regions. To avoid this inconvenience, we consider another representation of $U(x)$:
\begin{align} \label{eq:transformationrepresentation2}
 U(x) = \left \{ \begin{array}{lll}
 U_0 (x, \minus \frac{L}{2}) S_- &\quad \text{for}\,\,\,\minus\frac{L}{2} < x < \minus \frac{\epsilon}{2} \,,
 \\ [1mm] S_0 e^{\hat{\omega} (p-\frac{1}{2} -\frac{x}{\epsilon})}  &\quad  \text{for}\,\,\,
|x| < \frac{\epsilon}{2} \,, \\ [1mm]
  U_0 (x, \frac{L}{2}) S_+ &\quad  \text{for}\,\,\,
\frac{\epsilon}{2} < x < \frac{L}{2} \,,
 \end{array} \right.
\end{align}
with the decompositions $U_0 (\minus \epsilon/2, \minus L/2) = S_0 e^{p \hat{\omega}} S_-^{-1}$ and $U_0 (L/2, \epsilon/2) = S_+ e^{ (1-p) \hat{\omega}} S_0^{-1}$. Here, $0 \leq p \leq 1$, $S_0$ is a $\hat{z}$ rotation,  and $\epsilon>0$ is an infinitesimal number to be taken to zero below. Notably, $U(L/2)= S_+$ and $U(\minus L/2)= S_-$ in this representation, which keeps the Lagrangian in the lead and QPC regions unaffected. At the same time, the effect of disorder is accumulated at the single point $x = 0$. Indeed, with the transformation $\tilde{\psi} (x) = U^T (x) \psi (x)$ and Eq.~\eqref{eq:transformationrepresentation2}, Eq.~\eqref{eq:armso3} becomes
\begin{align} \label{eq:armso3rotatedbasis_new}
   \mathcal{L}_n &=  \int_{x \in \mathcal{R}_{\text{Arm}}} dx \big [\minus i \tilde{\psi}^T (\partial_t - \bar{v}_n \partial_x) \tilde{\psi} \big ] \nonumber \\ 
 & \quad \quad \quad \quad  - \frac{2 i \theta \bar{v}_n}{\sqrt{\pi a}}
   \sin (\phi_n (x=0)) \chi (x = 0)\,.
\end{align}
We now see that the action for the entire system has two fixed points, which correspond to the two values of $\theta=0,\pi$ where disorder can be fully removed. While $\theta = 0$ directly removes the disorder in Eq.~\eqref{eq:armso3rotatedbasis_new}, for $\theta = \pi$, the term $e^{\hat{\omega}} = \text{diag} (1, -1, -1)$ in Eq.~\eqref{eq:totalphasefactor}. The two minus terms are fully accounted for upon substituting $\phi_n \rightarrow -\phi_n$ and $\chi \rightarrow -\chi$ in the region $x > L/2$. Equivalently, in a picture of a Bloch sphere on which $\psi$ rotates by evolving with $U(x)$, $\psi$ does not rotate at all for $\theta = 0$ (clean limit), while $\psi$ rotates from the north to the south pole or vice versa for $\theta = \pi$. A more detailed RG analysis for the entire device shows that $\theta = 0$ and $\theta = \pi$ indeed correspond to a stable and unstable fixed point, respectively~\cite{SuppMat}.

With the two possible fixed points at hand, we determine the coherent two-terminal conductance $G$ in Fig.~\ref{fig:APF_Setup}. To this end, we boundary match $\phi_n$ and $\phi_c$ on all interfaces between lead and arm regions as well as between QPC regions and arm regions~\cite{SuppMat}. For all possible combinations of arm fixed points (i.e., for $\theta_{ij} = 0$ or $\theta_{ij}=\pi$ with labels $i\in\{$L\,(left), R\,(right)$\}$ and $j\in\{$U\,(up), D\,(down)$\}$), we find 
\begin{align} \label{eq:conductancecoh}
 G = \left \{ \begin{array}{lll}
 2 + \frac{1}{17} & \,\, \text{for}\,\, \text{all}\,\, \theta_{ij} = \pi \,,
 \\ [1mm] 2 + \frac{1}{9} &\,\,   \text{for} \ 
 \theta_{\rm LU} = \theta_{\rm LD} = \pi \ \text{and} \  \theta_{\rm RU} \cdot \theta_{\rm RD} = 0 \,, \\
 & \,\,\,   \text{or} \ 
 \theta_{\rm RU} = \theta_{\rm RD} = \pi \ \text{and} \  \theta_{\rm LU} \cdot \theta_{\rm LD} = 0 \,,
 \\  [1mm]
 2 + 1  &\,\,  \text{otherwise}\,. 
 \end{array} \right .
\end{align}
Here, 2 represents the contribution of the two lowest-Landau-level integer modes.
 At sufficiently low $T$, such that $L< L_T \equiv \hbar v_n / T$, each arm region is separately driven to the $\theta_{ij}= 0$ fixed point. In this clean device regime, the conductance is maximal at $G=3$. In contrast, for $L_T < L < \ell_{\text{eq}}^c$, $G$ fluctuates and its value depends on the specific disorder realizations, i.e., the precise values of all $\theta_{ij}$. These fluctuations generically range between $G = 2 + 1/ 17$ ($\theta_{ij} = \pi$) and $G = 3$. As $L$ exceeds ${\ell^c_{\text{eq}}}$ (i.e., the incoherent regime), $G$ approaches $G = 2 + 1/3$ as discussed above. 

We emphasize that, due to its unique hole-conjugate nature, the aPf edge permits upstream charge transport, allowing $G > \nu_B = 5/2$ in the coherent regime. By contrast, the Pf and phPf edges do not entail upstream charge transport (assuming no edge reconstruction), so that $G$ cannot exceed $5/2$. Hence, $G > 5/2$ in the QPC geometry is a hallmark of the aPf state. 


\textcolor{blue}{\noindent{\textit{Summary.---}}}We studied $\nu=5/2$ edge transport in a QPC device. In the incoherent regime, we found that among the Pf, phPf, and aPf candidate states, the aPf uniquely permits \textit{two} mechanisms that generate a conductance plateau at $G=7/3$. We proposed that shot noise, $S$, differentiate these mechanisms: (i) For $\nu_{\rm QPC}=3$, which is realized only for the aPf state,
$S$ reaches a maximum value $S \approx 0.12 e^3 V / h$ in the experimentally relevant regime~\eqref{eq:lengthscales}. (ii) For $\nu_{\rm QPC}=7/3$, $S\propto \sqrt{\ell_{\text{eq}}^h / L_{\text{QPC}}}$. We also studied the conductance in the coherent regime, where $G>5/2$ emerges uniquely for the aPf state, thus providing another fingerprint. Our results will enhance the prospects to experimentally pin-point the $\nu=5/2$ state in GaAs/AlGaAs, graphene, and further 2D materials. Our approach can be adapted to investigate other even-denominator FQH states. 

\textcolor{blue}{\noindent{\textit{Acknowledgments.---}}} 
J.P. and A.D.M. acknowledge support by the DFG Grant MI 658/10-2 and the German-Israeli Foundation Grant I-1505-303.10/2019. This project has received funding from the European Union’s Horizon 2020 research and innovation programme under grant agreement No. 101031655 (TEAPOT). 

\bibliography{References.bib}

\clearpage
\newpage
\onecolumngrid
\global\long\def\thesection{S\Alph{section}}
\global\long\def\thesubsection{\Roman{subsection}}
\setcounter{equation}{0}
\setcounter{figure}{0}
\setcounter{table}{0}
\setcounter{page}{1}
\renewcommand{\theequation}{S\arabic{equation}}
\renewcommand{\thefigure}{S\arabic{figure}}
\renewcommand{\bibnumfmt}[1]{[S#1]}
\renewcommand{\citenumfont}[1]{S#1}

\bigskip
\begin{center}
\large{\bf Supplemental Material for "Fingerprints of anti-Pfaffian topological order in quantum point contact transport"}\\
\end{center}
\begin{center}
Jinhong Park$^{1,2}$, Christian Sp\r{a}nsl\"{a}tt$^{3}$, and Alexander D. Mirlin$^{1,2}$
\\
{\it $^{1}$Institute for Quantum Materials and Technologies, Karlsruhe Institute of Technology, 76021 Karlsruhe, Germany\\$^{2}$nstitut f{\"u}r Theorie der Kondensierten Materie, Karlsruhe Institute of Technology, 76128 Karlsruhe, Germany\\$^{3}$Department of Microtechnology and Nanoscience (MC2), Chalmers University of Technology, S-412 96 G\"oteborg, Sweden}\\
(Dated: \today)
\end{center}

This supplemental material contains 7 sections. Section~\ref{sec:Incoherent} summarizes key features of incoherent transport along fractional quantum Hall (FQH) edges, as studied previously in Refs.~\cite{Nosiglia2018, Protopopov2017, Park2019,Spanslatt2019}. Based on the analysis in Sec.~\ref{sec:Incoherent}, we derive in Sec.~\ref{sec:conductanceplateau} the conductance plateau formula~\eqref{eq:nontrivialconductanceplateau} in the main text. In Sec.~\ref{sec:noiseplateau}, we investigate in detail the ``on-plateau'' shot noise characteristics for the anti-Pfaffian (aPf) state. Sections \ref{sec:RGanalysis} and \ref{sec:two-terminalconductance} address the coherent regime. Specifically, in Sec.~\ref{sec:RGanalysis} we perform a renormalization group (RG) analysis near two fixed points for the aPf edge attached to leads, and Sec.~\ref{sec:two-terminalconductance} contains our calculations of the two-terminal conductance in the quantum point contact (QPC) configuration depicted in Fig.~\ref{fig:APF_Setup} in the main text. Section~\ref{sec:conductancerange} shows that the two-terminal conductance in the QPC geometry ranges from $G = 2+1/17$ to $2+1$ (in units of $e^2/h$). Finally, in Sec.~\ref{sec:kuboformula}, we derive the Kubo formula used in our calculation of the two-terminal conductance.

\section{Recap: Incoherent transport in a line junction} 

\label{sec:Incoherent}

In this section, we recap the features of the incoherent transport regime of FQH edge modes, previously investigated in Refs.~\cite{Nosiglia2018, Protopopov2017, Park2019,Spanslatt2019,Hein2023}. 
This section presents the framework for the calculation of the conductance on QPC plateaus (part ``QPC conductance plateaus" of the main text and Sec.~\ref{sec:conductanceplateau} of this Supplemental Material) and details of derivation of a general formula for the nonequilibrium noise (part ``Shot noise in the incoherent regime'' of the main text). The latter result is used in 
Sec.~\ref{sec:noiseplateau} to calculate the noise on $G = 7/3$ plateaus for $\nu_B=5/2$. 


To describe incoherent transport along a FQH edge segment, we model it as a line junction consisting of multiple counter-propagating edge modes. Edge mode $i$ is associated with the filling factor discontinuities $\delta \nu_i$; The parameter $n_d$ ($1 \leq i \leq n_d$) specifies the number of modes propagating in the downstream direction and $n_u$ is the number of modes ($n_d + 1 \leq i \leq n_d + n_u$) propagating upstream; The inequality $\nu_d > \nu_u$ is required for the charge transport to propagate along the the downstream direction. The edge modes generically have different central charges $c_i$ which determines the amount of heat carried by each mode. In the incoherent regime, we can define a local, effective electrochemical potential (related to a local voltage) and a local, effective temperature for each individual mode. These local quantities are determined by solving a set of transport equations defined below in Eq.~\eqref{eq:voltagediffeq} and~\eqref{eq:energycurrentprofile}. 

\begin{figure}[t!]
\includegraphics[width =0.6\columnwidth]{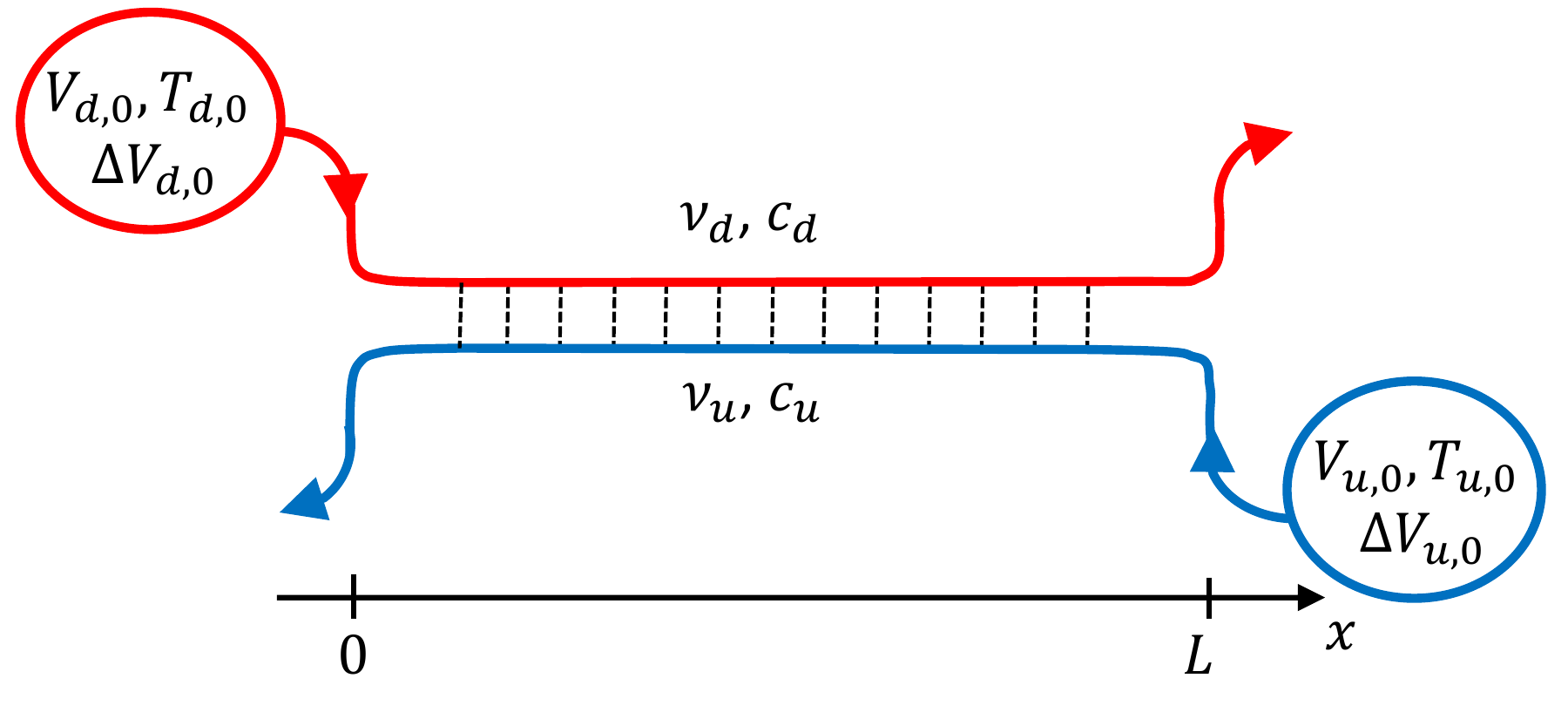}
\caption{{\bf Line junction setup} with length $L$ consisting of a downstream mode (in red, solid line) associated with the filling factor discontinuity $\nu_d$ and central charge $c_d$, and a upstream mode (blue, solid line) with the filling factor discontinuity $\nu_u$ and central charge $c_u$. The downstream and upstream mode emanate from the respective source with voltage $V_{d,0}$, temperature $T_{d,0}$, and the voltage fluctuation $\Delta V_{d, 0}$, and $V_{u,0}$, $T_{u,0}$, and $\Delta V_{u, 0}$, respectively. To describe equilibration between the modes, tunneling junctions (black, dashed lines) are introduced. 
}
\label{supFig:linejunction}
\end{figure}

To simplify the problem, we next assume that modes propagating in the same direction share the same local voltage and temperature. This approximation allows us to effectively reduce the full number of modes into only two effective, hydrodynamic modes propagating in opposite directions: One downstream mode associated with the total filling factor $\nu_d = \sum_{i = 1}^{n_d} \nu_i$ and the central charge $c_d = \sum_{i = 1}^{n_d} c_i $, and one upstream mode with $\nu_u = \sum_{i = n_d + 1}^{n_d + n_u} \nu_i$ and $c_u = \sum_{i = n_d + 1}^{n_d + n_u} c_i$, see Fig.~\ref{supFig:linejunction}. This approximation is not generically not valid, but it is excellent for the aPf edge structure since the Majorana mode and the bosonic mode $\delta \nu = \minus 1/2$ are geometrically close to each other and propagate in the same direction; For the aPf edge, we have $n_d = 1$, $c_d = 1$, $n_u = 1/2$, and $c_u=1+1/2 = 3/2$. Note that the $\delta \nu = \minus 1/2$ mode and the Majorana mode contribute $1$ and $1/2$ to $c_u$, respectively. 

To describe equilibration between the two effective modes above, we consider an array of $N$ tunneling junctions ($1 \leq j \leq N$). The total length of the line junction is $L \equiv N a$ with $a$ the distance between each adjacent tunneling junctions. Enforcing current conservation in each individual tunneling junction we have
\begin{align} \label{eq:currentconservation}
    I_{d, j+1} = I_{d, j} - I_{\text{tun},j}\,, \quad \quad 
    I_{u, j+1} = I_{u, j} + I_{\text{tun},j}\,,
\end{align}
where the tunneling current $I_{\text{tun},j}$ in tunneling junction $j$ takes the form
\begin{align} \label{eq:tunnelingcurrent}
    I_{\text{tun}, j} = g_j^c \frac{e^2}{h} (V_{d,j} - V_{u, j+1})\,.
\end{align}
Here, $V_{d, j}$ and $V_{u, j}$ are the effective voltages of the modes along the $j$th edge segment and $g_j^c$ describes the effective transmission across tunneling junction $j$. We neglect the non-trivial temperature dependence of $g_j^c$ from the Luttinger liquid description of the modes and we treat $g_j^c$ as an input parameter of the model. The voltages of the respective modes are related to the electric current carried by the modes as 
\begin{align}  \label{eq:currentvoltagerelation}
 I_{d, j} = \nu_d \frac{e^2}{h} V_{d, j}\,, \quad \quad   I_{u, j} = -\nu_u \frac{e^2}{h} V_{u, j}\,.
\end{align}
Combining Eqs.~\eqref{eq:currentconservation}-\eqref{eq:currentvoltagerelation} to linear order in $g^c_j$, as well as assuming small $g_j^c \ll 1$, we obtain
\begin{align}
\begin{pmatrix} 
     V_{d, j+1} \\ V_{u, j+1} 
\end{pmatrix} = \begin{pmatrix} 
     V_{d, j} \\ V_{u, j} 
\end{pmatrix} + g_j M_j \begin{pmatrix}
     V_{d, j} \\ V_{u, j}
\end{pmatrix}\,\,\,\, \text{with}\,\,M_j = \begin{pmatrix}
     -1/\nu_d & 1/\nu_d \\ -1/\nu_u & 1/\nu_u
\end{pmatrix}\,.
\end{align}
In the continuum limit, $N \rightarrow \infty$ and $a \rightarrow 0$, with $L=Na$ finite, we obtain the differential equations
\begin{align} \label{eq:voltagediffeq}
    \frac{d}{dx} \begin{pmatrix} 
     V_{d} (x) \\ V_{u} (x) 
\end{pmatrix} = \frac{1}{\ell_{\text{eq}}^{c}} M_j \begin{pmatrix} 
     V_{d} (x) \\ V_{u} (x) 
\end{pmatrix}\,.
\end{align}
Here, we assumed uniform tunneling rates $g_j^c = g^c$ and $\ell_{\text{eq}}^c$ is defined as $\ell_{\text{eq}}^c \equiv a/g^c$. Note that the charge equilibration between the edge modes is fully characterized by the single length scale $\ell_{\text{eq}}^c$. The differential equations \eqref{eq:voltagediffeq} can be  solved analytically with the boundary conditions $V_{d} (0) = V_{d, 0}$ and $V_{u} (L) = V_{u, 0}$, which leads to the following two voltage profiles along the edge:
\begin{align} \label{eq:voltageprofile}
V_{d} (x) = \frac{\nu_d V_{d, 0} - \nu_u (V_{d, 0} - V_{u, 0})e^{-(L-x)/(\ell_{\text{eq}}^c \tilde{\nu})} - \nu_u V_{u, 0} e^{-L/{(\ell_{\text{eq}}^c \tilde{\nu})}}}{\nu_d - \nu_u e^{-L/(\ell_{\text{eq}}^c \tilde{\nu})}}\,, \nonumber \\ 
V_{u} (x) = \frac{\nu_d V_{d, 0} - \nu_d (V_{d, 0} - V_{u, 0})e^{-(L-x)/(\ell_{\text{eq}}^c \tilde{\nu})} - \nu_u V_{u, 0} e^{-L/(\ell_{\text{eq}}^c \tilde{\nu} )}}{\nu_d - \nu_u e^{-L / (\ell_{\text{eq}}^c \tilde{\nu})}}\,.
\end{align}
Here, the parameter $\tilde{\nu} \equiv (1/\nu_u - 1/\nu_d)^{-1}>0$ since $\nu_d > \nu_u$. Then, the voltage of the outgoing modes is given by
\begin{align} \label{eq:voltageoutgoingmodes}
   V_{d} (x = L) &= \frac{(\nu_d - \nu_u ) V_{d, 0} + \nu_u V_{u, 0} - \nu_u V_{u, 0} e^{-L/{(\ell_{\text{eq}}^c \tilde{\nu})}}}{\nu_d - \nu_u e^{-L/(\ell_{\text{eq}}^c \tilde{\nu})}} \xrightarrow{\ell_{\text{eq}}^c \ll L} \frac{(\nu_d - \nu_u ) V_{d, 0} + \nu_u V_{u, 0}}{\nu_d} \,, \nonumber \\ 
V_{u} (x = 0) &= \frac{\nu_d V_{d, 0} - (\nu_d V_{d, 0} - (\nu_d - \nu_u ) V_{u, 0})e^{-L/(\ell_{\text{eq}}^c \tilde{\nu})}}{\nu_d - \nu_u e^{-L / (\ell_{\text{eq}}^c \tilde{\nu})}} \xrightarrow{\ell_{\text{eq}}^c \ll L}  V_{d,0}\,.
\end{align}
From Eq.~\eqref{eq:voltageoutgoingmodes}, one obtains the downstream conductance $G_d$ with $V_{u, 0} = 0$ and the upstream conductance $G_u$ with $V_{d,0} = 0$, respectively, as
\begin{align} \label{eq:downupcond}
    G_d = \frac{I_{d} (x = L)}{V_{d, 0}}  =  \frac{\nu_{d} e^2}{h} \frac{V_d(x = L)}{V_{d, 0}} = (\nu_d - \nu_u) \frac{e^2}{h} \,, \quad \quad
    G_u = \frac{I_{u} (x = 0)}{V_{u, 0}}  = 0\,.
\end{align}
The absence of upstream charge transport, $G_u = 0$, is a key result in the incoherent regime. 

With the same approach, we obtain the two edge mode temperature profiles. Energy conservation in each individual tunneling junction gives
\begin{align} \label{eq:energyconservation}
    J_{d, j+1} = J_{d, j} - J_{\text{tun},j}\,, \quad \quad
    J_{u, j+1} = J_{u, j} + J_{\text{tun},j}\,,
\end{align}
where the local energy current carried by each mode is related to the local temperatures and voltages as
\begin{align} \label{eq:relationenergycurrent}
    J_{d, j} = \frac{\nu_d e^2}{2h} V_{d, j}^2 + \frac{c_d}{2} \kappa_0 T_{d, j}^2\,, \quad \quad
    J_{u, j} = -\frac{\nu_u e^2}{2h} V_{u, j}^2 -\frac{c_u}{2} \kappa_0 T_{u, j}^2\,.
\end{align}
Further, the tunneling energy current $J_{\text{tun}, j}$ reads
\begin{align} \label{eq:tunnelingenergycurrent}
    J_{\text{tun}, j} = g_j^c \frac{e^2}{2 h} (V_{d,j}^2 - V_{u, j+1}^2) + g_j^h \frac{\kappa_0}{2} (T_{d,j}^2 - T_{u, j+1}^2)\,,
\end{align}
in terms of the two effective, local temperatures $T_{d,j}$ and $T_{u,j}$. The parameter $\kappa_0 \equiv \frac{\pi^2 k_B^2}{3h}$ and $k_B$ is the Boltzmann constant. The tunneling energy current \eqref{eq:tunnelingenergycurrent} has two contributions: The first term is associated with the charge tunneling processes and thus the 
coefficient of this term is given by $g_j^c$. By contrast, the second term originates from the temperature difference between the edge modes. As a dimensionless coefficient of this second term, we introduce another parameter $g_j^h$. 
Combining Eqs.~\eqref{eq:energyconservation}-\eqref{eq:relationenergycurrent} and taking the continuum limit, $N \rightarrow \infty$ and $a \rightarrow 0$ with $L=Na$ finite, we obtain the differential equations for the local heat currents $J_{d/u}^h(x) \equiv \pm c_{d/u} \kappa_0 T_{d/u}^2 (x) /2$ as
\begin{align} \label{eq:energycurrentprofile}
    \frac{d}{dx} \begin{pmatrix} 
     J_{d}^h (x) \\ J_{u}^h  (x) 
\end{pmatrix} = \frac{1}{\ell_{\text{eq}}^{h}} \begin{pmatrix}
     -1/c_d & -1/c_u \\ 1/c_d & 1/c_u
\end{pmatrix} \begin{pmatrix} 
     J_{d}^h  (x) \\ J_{u}^h  (x) 
\end{pmatrix}+ \frac{e^2(V_{d} (x)- V_{u} (x))^2}{ h \ell_{\text{eq}}^c}
\begin{pmatrix} 
    1/2 \\ 1/2
\end{pmatrix}\,.
\end{align}
The heat equilibration length $\ell_{\text{eq}}^h$ is defined here as $\ell_{\text{eq}}^h \equiv a/g_{j}^h = a/g^h$. 

The parameters $g_j^c$ and $g^h_j$ are in general different, implying a violation of the Wiedemann-Franz law~\cite{Kane1996, Nosiglia2018} related to the fact the FQH edges originate from a strongly interacting system. Furthermore, there is in general the interaction between the edge modes, which was discarded in the above discussion. This leads to renormalization of $\ell_{\text{eq}}^c$, making $\ell_{\text{eq}}^c \ll \ell_{\text{eq}}^h$ in a range of strong interactions~\cite{Srivastav2021May}.  Recent experiments showed that this is indeed the case
~\cite{Srivastav2021May,Srivastav2022Sep,Melcer2022Jan, Dutta2022Sep}. Remarkably, it was found that
$\ell_{\text{eq}}^h$ can be two orders of magnitude larger than $\ell_{\text{eq}}^c$.
Below, we treat the heat and charge equilibration lengths, $\ell_{\text{eq}}^h$ and $\ell_{\text{eq}}^c$, as two independent parameters satisfying $\ell_{\text{eq}}^c \ll \ell_{\text{eq}}^h$.

The second term in Eq.~\eqref{eq:energycurrentprofile} describes the Joule-heating contribution generated by the voltage difference across the junctions. The total Joule-heating power, $P_j$, is obtained by integrating this second term over the whole edge using the voltage profiles \eqref{eq:voltageprofile}
\begin{align} \label{eq:jouleheating}
    P_{J} = \frac{e^2}{h \ell_{\text{eq}}^c}\int_{0}^{L} dx (V_{d} (x) - V_{u} (x))^2 \xrightarrow{\ell_{\text{eq}}^c \ll L} 
    \frac{(\nu_d - \nu_u) \nu_u e^2 (V_{d, 0} - V_{u, 0})^2}{2 \nu_d}\,.
    \end{align}

By solving the differential equations Eq.~\eqref{eq:energycurrentprofile} with the boundary conditions $T_d (x = 0) = T_{d, 0}$, $T_{u} (x = L) = T_{u,0}$, $V_d (x=0) = V_{d, 0}$, and $V_u (x =L) = V_{u, 0}$, we obtain the temperatures of the outgoing modes, $T_d (x = L)$ and $T_{u} (x = 0)$.
We analytically compute the temperatures of the outgoing modes for two different length hierarchy limits: (i) The fully thermal equilibrated regime for which $\ell_{\text{eq}}^c \ll \ell_{\text{eq}}^h \ll L$ and (ii) the absent thermal equilibration regime, where $\ell_{\text{eq}}^c \ll L \ll \ell_{\text{eq}}^h$. We also assume $c_d < c_u$, as appropriate for the aPf edge structure. In the fully thermal equilibrated regime (i), we obtain 
\begin{align} \label{eq:temperatureapprox}
    T_{u}^2 (x = 0) &= \frac{c_u - c_d}{c_u} T_{u, 0}^2 + \frac{c_d}{c_u} T_{d, 0}^2 + \frac{(c_u - c_d) }{c_u^2 L_0} P_J \,, \nonumber \\ 
    T_{d}^2 (x = L) &= T_{u, 0}^2 +\frac{(c_d + c_u)}{c_u c_d  L_0}P_J\,,
\end{align}
where $L_0 \equiv \pi^2 k_B^2 /( 3 e^2) = \kappa_0 h / e^2$ is the Lorenz number and $P_J$ is given in Eq.~\eqref{eq:jouleheating}. In the absent thermal equilibration regime (ii), we instead obtain 
\begin{align} \label{eq:temperatureapprox2}
    T_{u}^2 (x = 0) = T_{u, 0}^2 + \frac{P_J }{c_u L_0}  \,, \quad \quad
    T_{d}^2 (x = L) = T_{d, 0}^2 +\frac{P_J}{c_d L_0}\,.
\end{align}
Equations \eqref{eq:temperatureapprox} and \eqref{eq:temperatureapprox2} in the two distinct thermal equilbration regimes will be used in Sec.~\ref{sec:noiseplateau} to obtain the analytic formulas for the noise in the two regimes. 

With voltage and temperature profiles at hand, we next compute the electrical noise generated in the line junction. Current conservation at each tunneling junction results in the following relations for the electrical current fluctuations
\begin{align} \label{eq:currentconservationfluc}
    \Delta I_{d, j+1} =  \Delta I_{d, j} - \Delta I_{\text{tun},j}\,, \quad \quad
    \Delta I_{u, j+1} = \Delta I_{u, j} + \Delta I_{\text{tun},j}\,,
\end{align}
where the fluctuations in the inter-mode tunneling currents are given by
\begin{align} \label{eq:tunnelingcurrentfluc}
    \Delta I_{\text{tun},j} &=   g_j^c \frac{e^2}{h} (\Delta V_{d,j} - \Delta V_{u, j+1}) + \delta I_{\text{tun}, j}\,. 
\end{align}
The first term in this expression is the extrinsic current fluctuations originating from the voltage fluctuations impinging on the $j$th tunneling junction. The second term is an intrinsic thermal contribution from the junction itself. In the incoherent regime, the intrinsic current fluctuations generated at different tunneling junctions are completely independent of each other, and we may write
\begin{align}
    \overline{\delta I_{\text{tun}, j} \delta I_{\text{tun}, j'}} \approx 2g^c_j \frac{e^2}{h} \delta_{jj'}\left[k_B T_{d,j}+k_B T_{u,j}\right],
\end{align}
whenever charge equilibration is efficient, $V_{d,j}\approx V_{u,j}$. The effect of current fluctuations along the edge modes is directly reflected in the voltage fluctuation as
\begin{align} \label{eq:voltagecurrentrelation}
   \Delta I_{d/u, j} = -\frac{e^2 \nu_{d/u}}{h}\Delta V_{d/u, j}\,. 
\end{align}

Combining Eqs.~\eqref{eq:currentconservationfluc}-\eqref{eq:voltagecurrentrelation} and taking the continuum limit, we arrive at the following equation for the total current fluctuations along the edge:
\begin{align} \label{supeq:noiseformula}
  \Delta I \equiv \Delta I_{d} (x) - \Delta I_{u} (x) = G_d \Delta V_{d, 0} - G_u \Delta V_{u, 0} + \delta I_{\text{line}}\,, \nonumber \\ 
  \text{with}\, \quad  \delta I_{\text{line}} \equiv \frac{1}{a}\int_{0}^{L} dx \delta I_{\text{tun}} (x) \frac{(\nu_d - \nu_u) e^{-x/ (\ell_{\text{eq}}^c  \tilde{\nu})}}{\nu_d - \nu_u e^{-L/ (\ell_{\text{eq}}^c \tilde{\nu})}}\,.
\end{align}
Here, the boundary conditions of the incoming modes are $\Delta V_{d} (x = 0) = \Delta V_{d, 0}$ and $\Delta V_{u} (x = L) = \Delta V_{u, 0}$, and the downstream and upstream conductances $G_d$ and $G_u$ are given in Eq.~\eqref{eq:downupcond}. Note that $\Delta I$ does not depend on $x$ due to current conservation. While the first term in Eq.~\eqref{supeq:noiseformula} is the extrinsic contribution from the current fluctuations of the incoming modes, the second term, $\delta I_{\text{line}}$, is the contribution from the intrinsic current fluctuations generated along the line junction. Assuming noiseless input voltages, $\Delta V_{d, 0} = \Delta V_{u, 0} = 0$, we arrive at the following analytical expression for the noise
\begin{align}
 S_{\text{line}} &= \overline{(\Delta I)^2} = \frac{2 (\nu_d - \nu_u)^2}{\ell_{\text{eq}}^c} \frac{e^2}{h} \int_{0}^{L} dx \left [k_B (T_{d} (x) + T_{u} (x)) \right]
 \frac{e^{-2x/(\ell_{\text{eq}}^c \tilde{\nu})}}{(\nu_d - \nu_u e^{-L/ (\ell_{\text{eq}}^c \tilde{\nu})})^2} 
 \nonumber \\ &\xrightarrow{\ell_{\text{eq}}^c \ll \ell_{\text{eq}}^h} \frac{e^2}{h}\frac{(\nu_d - \nu_u) \nu_u}{\nu_d} k_B(T_{d} (x = 0) + T_{u} (x = 0)),
\end{align}
which is Eq.~\eqref{eq:noiselinejunction} in the main text.

\section{Derivation of conductance plateaus for $\nu_{\text{QPC}} > \nu_B$} \label{sec:conductanceplateau}
\begin{figure}[t!]
\includegraphics[width =0.6\columnwidth]{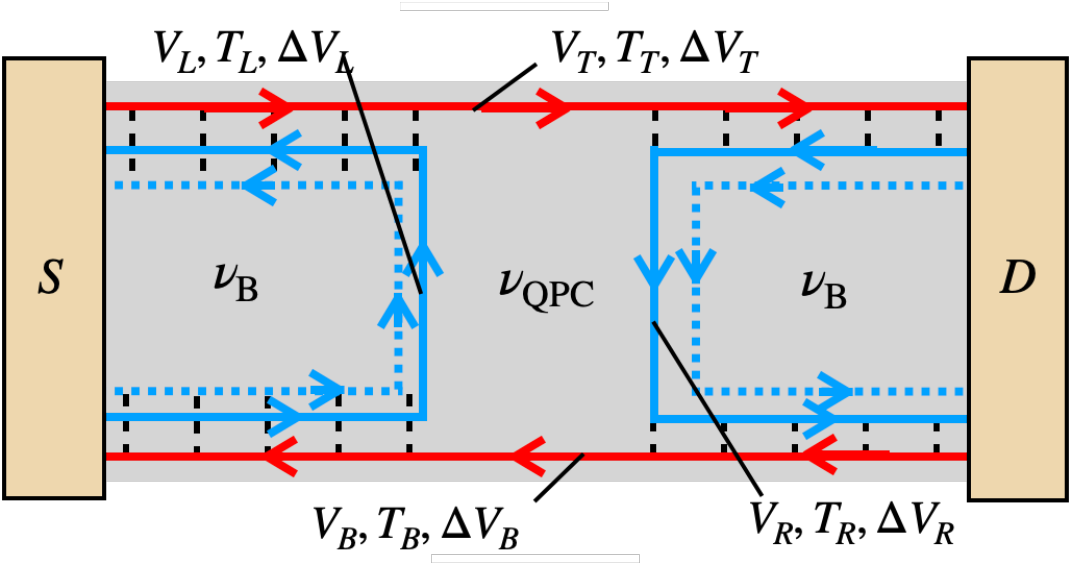}
\caption{{\bf QPC setup} for the aPf state with the formation of a local region $ \nu_{\text{QPC}} > \nu_B$.
The locations of voltages, temperatures, and the fluctuation of voltages, referred to in Secs.~\ref{sec:conductanceplateau} and \ref{sec:noiseplateau}, are marked out.
}
\label{supFig:qpc}
\end{figure}

In this section, we derive the conductance formula~\eqref{eq:nontrivialconductanceplateau} in the main text. We consider the formation of a local region $\nu_{\rm QPC}>\nu_B$, see Fig.~\ref{supFig:qpc}. Modes associated with the filling factor $\nu_T$ are fully transmitted (not shown in Fig.~\ref{supFig:qpc})  and we assume that they have zero coupling to the other modes. 

Let us first consider the case where $\nu_T = 0$. The geometry of the arms of the QPC is essentially identical with the line junction setup (Fig.~\ref{supFig:linejunction}) considered in Sec.~\ref{sec:Incoherent}, upon identification $\nu_d = \nu_{\text{QPC}}$ and $\nu_{u} = \nu_{\text{QPC}} - \nu_B$. In the incoherent regime $L \gg \ell_{\text{eq}}^c$, we can thus use all the formulas obtained in Sec.~\ref{sec:Incoherent}. The voltage on the source contact is $V_S=V$ and we take the drain to be grounded $V_D=0$. From Eq.~\eqref{eq:voltageoutgoingmodes}, we then obtain the following relations between the local voltages in the device: $V_T$, $V_B$, $V_R$, and $V_L$ (see Fig.~\ref{supFig:qpc}): 
\begin{align}
    V_T = \frac{\nu_B V + (\nu_{\text{QPC}}-\nu_B) V_L}{\nu_{\text{QPC}}}\,, \quad \quad V_T = V_R\,, \quad \quad 
    V_B =  \frac{(\nu_{\text{QPC}}-\nu_B) V_R}{\nu_{\text{QPC}}}\,, \quad \quad V_B = V_L\,.
\end{align}
From these expressions we further find
\begin{align} \label{supeq:voltages}
    V_T = V_R = \frac{\nu_{\text{QPC}}}{2\nu_{\text{QPC}} - \nu_B }V\,, \quad V_B = V_L = \frac{\nu_{\text{QPC}} - \nu_B }{2\nu_{\text{QPC}} - \nu_B}V\,.
\end{align}
Then, the two-terminal conductance is given by 
\begin{align}
\label{eq:G_formula_SuppMat}
    G \equiv \frac{I_D}{V} =\frac{e^2}{h} \frac{\nu_B V_T}{V} =\frac{e^2}{h} \frac{\nu_{\text{QPC}} \nu_B}{2\nu_{\text{QPC}} - \nu_B }\,. 
\end{align}
This result for $G$ with $\nu_T = 0$ can be straightforwardly generalized to finite $\nu_T$ by (i)  the replacements $\nu_B \rightarrow \nu_B - \nu_{T}$ and $\nu_{\text{QPC}} \rightarrow \nu_{\text{QPC}} - \nu_T$ and (ii) the addition of the direct contribution $\nu_T e^2 /h$. Applying (i) and (ii) to Eq.~\eqref{eq:G_formula_SuppMat} leads directly to Eq.~\eqref{eq:nontrivialconductanceplateau} in the main text. 

\section{Noise on the conductance plateaus} \label{sec:noiseplateau}

In this section, we compute the noise on the conductance plateau $G = 7e^2/(3h)$ for the aPf edge in the QPC configuration with $\nu_{\text{QPC}} = 3 > \nu_B = 5/2$ (part ``Shot noise in the incoherent regime'' of the main text). We compute the noise analytically in two distinct limits for the degree of thermal equilibration: 
(i) The fully thermal equilibrated regime, where $\ell_{\text{eq}}^c \ll \ell_{\text{eq}}^h \ll L$, and (ii) the absent thermal equilibration regime for which $\ell_{\text{eq}}^c \ll L \ll \ell_{\text{eq}}^h$. 

The geometry of the arms of the QPC is essentially identical to the line junction setup in Fig.~\ref{supFig:linejunction} considered in Sec.~\ref{sec:Incoherent}, upon the identification $\nu_d = 1$, $\nu_u = 1/2$, $c_d = 1$, and $c_u =3/2$. 
We compute the voltages $V_T$, $V_B$, $V_R$, $V_L$ and temperatures $T_T$, $T_B$, $T_R$, $T_L$ in Fig.~\ref{supFig:qpc} by solving the differential equations~\eqref{eq:voltagediffeq} and~\eqref{eq:energycurrentprofile} for the electric and energy current, respectively, along each edge segment in the device. As boundary conditions for the two contacts we use $V_S = V$, $V_D = 0$ and temperatures $T_S=T_D =0$. For full charge equilibration, Eq.~\eqref{eq:voltageoutgoingmodes} gives us the voltages $V_T$, $V_B$, $V_R$, $V_L$ as
\begin{align}
    V_{T} = \frac{2V}{3}, \quad V_{R} = \frac{2V}{3}, \quad V_{B} = \frac{V}{3}, \quad V_{L} = \frac{V}{3}\,.
\end{align}
Let us now focus on regime (i), i.e., $\ell_{\text{eq}}^h \ll L$. By using the analytic formulas Eqs.~\eqref{eq:temperatureapprox} and \eqref{eq:jouleheating}, we obtain the following relations between the temperatures $T_U$, $T_D$, $T_R$, and $T_L$:
\begin{align}
   T_T^2 = T_L^2 + \frac{(c_d + c_u)}{18 c_d c_u L_0} V^2\,, \quad 
   T_R^2 = \frac{c_d}{c_u} T_U^2\,, \quad
   T_B^2 = T_R^2 +  \frac{(c_d + c_u)}{18 c_d c_u L_0} V^2\,, \quad 
   T_L^2 =\frac{c_d}{c_u} T_B^2\,,
\end{align}
which for $c_d = 1$ and $c_u = 3/2$ gives us
\begin{align} \label{eq:thermalequilibration1}
   T_T=T_B= \sqrt{\frac{5}{6}} \frac{eV}{\pi k_B}\,, \quad 
   T_R =T_L = \frac{\sqrt{5}}{3} \frac{eV}{\pi k_B}\,.
\end{align}
In regime (ii), i.e., for $\ell_{\text{eq}}^h \gg L$, Eqs.~\eqref{eq:jouleheating} and~\eqref{eq:temperatureapprox2} lead instead to 
\begin{align}
   T_T^2 = \frac{V^2}{18 c_d L_0} \,, \quad 
   T_B^2 =\frac{V^2}{18 c_d L_0}\,, \quad 
   T_L = T_R = 0\,,
\end{align}
whose solution is
\begin{align}
\label{eq:thermalequilibration2}
   T_T = T_B  = \frac{eV}{\sqrt{6}\pi k_B}\,, \quad 
   T_L = T_R = 0\,. 
\end{align}
That the temperatures~\eqref{eq:thermalequilibration2} are smaller than those in Eq.~\eqref{eq:thermalequilibration1} can be understood from the upstream, ballistic transport of the generated heat (from the QPC region to the leads) in regime (ii). This ballistic and upstream heat transport prevents heat to return back to the QPC region, thus resulting in lower temperatures.

We next compute the noise by applying Eq.~\eqref{supeq:noiseformula} to the arm regions by identifying $G_d = e^2 /(2h)$ and $G_u = 0$, and then connecting the arm regions by charge conservation. We take that the contact voltages do not fluctuate in time, i.e., we assume ideal voltage sources. Then, we obtain the following equations relating the voltage fluctuations at the different boundaries:
\begin{align} \label{eq:qpccurrentfluctuations}
 \Delta V_{T} =  \frac{\Delta V_{L}}{2}\,, &\quad  \Delta V_{B} =  \frac{\Delta V_{R}}{2}\,, \nonumber \\ 
   \frac{e^2}{2h} (\Delta V_{T} - \Delta V_{R}) = \delta I_{\text{line}, \text{RU}}\,, &\quad
    \frac{e^2}{2h}(\Delta V_{B} - \Delta V_{L}) =  \delta I_{\text{line}, \text{LD}}
\end{align}
Here, $\delta I_{\text{line}, \text{RU}/\text{LD}}$ denote the current fluctuation generated in the 
noise spots (the yellow regions in Fig.~\ref{fig:Noises} in the main text), located at the upper, right (lower, left) corner, respectively. 
Then, the total current fluctuations measured in the drain $D$ is given by 
\begin{align}
    \Delta I &= \frac{e^2}{h} (\Delta V_{T} - \Delta V_{B}) = \frac{2}{3} (\delta I_{\text{line}, \text{RU}} - \delta I_{\text{line}, \text{LD}})\,,
\end{align}
and thus the noise measured in $D$ becomes
\begin{align}
    S = \overline{(\Delta I)^2} = \frac{4}{9} (\overline{\delta I_{\text{line}, \text{RU}}^2}+\overline{\delta I_{\text{line}, \text{LD}}^2}) 
     = \frac{8}{9} S_{\text{NS}}\,.
\end{align}
In the final equality, we used $\overline{\delta I_{\text{line}, \text{RU}}^2} = \overline{\delta I_{\text{line}, \text{LD}}^2}$ by assuming that the geometry of our device is symmetric. The numerical coefficient $8/9$ is thus a conversion factor from the noise generated in a single noise spot, $S_{\rm NS}$ to the total noise $S$ measured in $D$, as discussed in the main text. Next, by using Eq.~\eqref{supeq:noiseformula}, we obtain the analytic results for $S$ in the two distinct thermally equilibrated regimes (i) and (ii). In the thermally equilibrated regime (i), we obtain the noise $S$ by using Eq.~\eqref{eq:thermalequilibration1} and find
\begin{align}
    S =  \frac{8}{9} S_{\text{NS}} \approx 0.12 \frac{e^3 V}{h}\,.
\end{align}
A corresponding Fano factor can be defined as
\begin{align}
\label{eq:Fano_Def}
    F \equiv \frac{S}{2 e I_\text{imp} T_{\text{eff}} (1- T_{\text{eff}})}\approx 0.53\,,
\end{align}
upon identification of the effective transmission $T_{\text{eff}} =  (1/3)/(1/2) = 2/3$ of the edge modes in the second Landau level, and the impinging current $I_{\text{imp}} = e^2V / (2h)$ in the incoherent regime. The definition~\eqref{eq:Fano_Def} has been used in several QPC experiments in the FQH regime, see e.g., Refs.~\cite{Bid2009shot,Bhattacharyya2019,Biswas2022shot}. By the same procedure as above, we obtain the noise and the corresponding Fano factor in the absent thermal equilibration regime (ii) as
\begin{align}
    S =  \frac{8}{9} S_{\text{NS}} \approx 0.029 \frac{e^3 V}{h}\,, \quad F \approx 0.13\,.
\end{align}
This value of $S$ is given as the green, dashed line in Fig.~\ref{Fig:noisechar} in the main text.
We stress that the Fano factor~\eqref{eq:Fano_Def} is unrelated to the charge of individually transmitted particles. Although the noise we have computed is proportional to the applied bias voltage $V$, similar to shot noise in tunneling experiments, our noise is of a very distinct origin: The application of $V$ heats the system to an effective temperature $\Delta T \sim V$, and this elevated temperature leads to noise $S \propto \Delta T \sim V$.

We finish this section by commenting on the noise on the conductance plateau produced for $\nu_B > \nu_{\text{QPC}}$. As shown in the main text, the noise on this plateau scales as $S \sim \sqrt{\ell_{\text{eq}}^h / L_{\text{QPC}}}$. This noise characteristic holds provided that $\ell_{\text{eq}}^h \gg L_{\text{QPC}}$, together with the emergence of localized modes inside the QPC region. Thus, such noise characteristics emerges regardless of the bulk topological order, including Abelian ones. We note that this result {\it does not} agree with Ref.~\cite{manna2023shot}, in which the noise approaches a constant value in the limit of no thermal equilibration, see~Eq.~(38) in Ref.~\cite{manna2023shot}. 

\section{ Coherent regime: RG analysis of a disordered aPf edge with leads} 

\label{sec:RGanalysis}

\begin{figure}[t!]
\includegraphics[width =0.5\columnwidth]{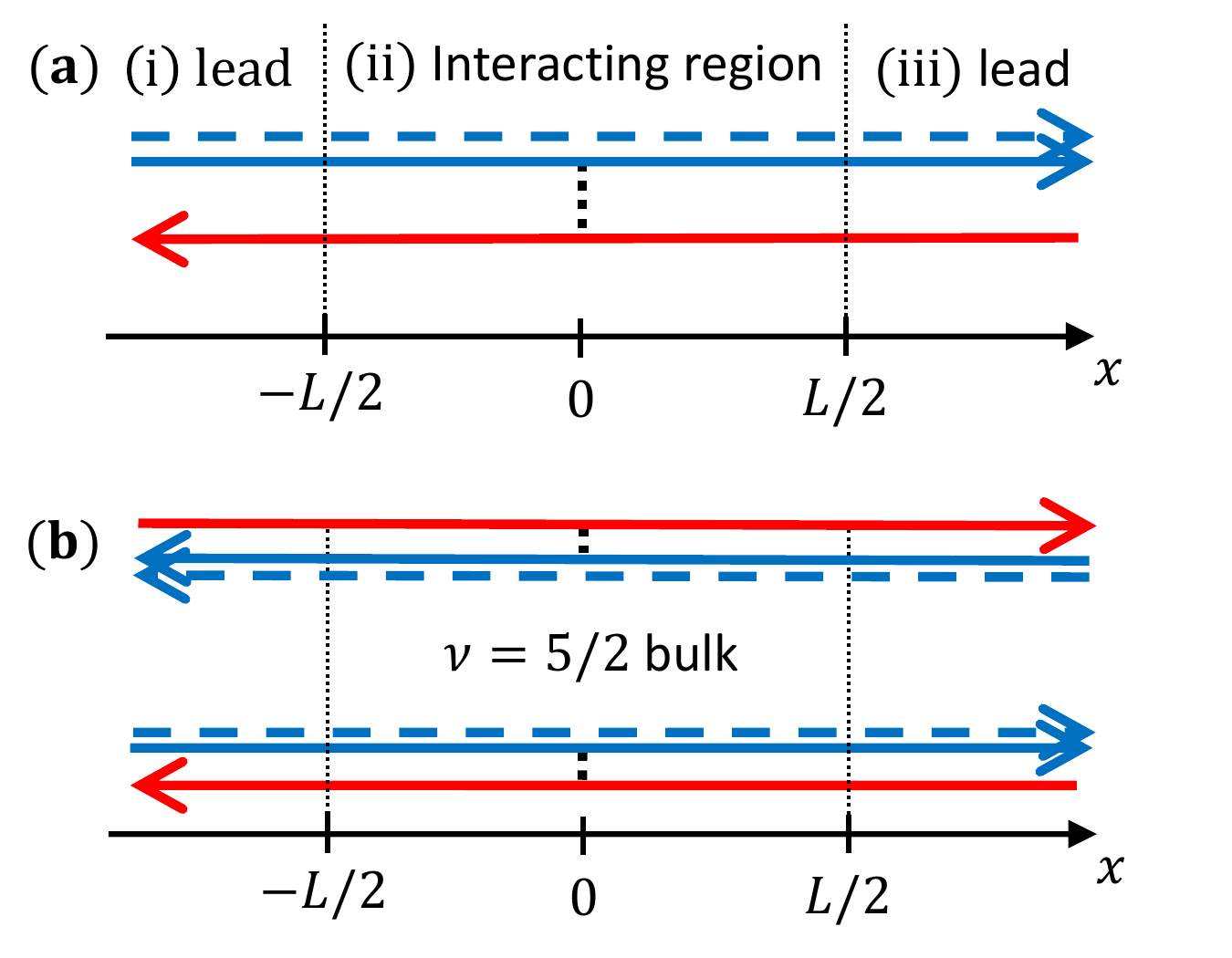}
\caption{{\bf Setups for the aPf state connected to leads}.  
{\bf (a)} The single edge consists of three regions: (i) the ``left lead'' $x < L/2$, (ii) the interacting and disordered region $|x| < L/2$, and (iii) the ``right lead'' $x > L/2$. Red and blue solid lines depict, respectively, $\delta \nu = 1$ and $\delta \nu = -1/2$ edge modes propagating in the opposite directions. The dashed blue line depicts a Majorana mode propagating in the same direction as the $\delta \nu = -1/2$ mode. {\bf (b)} Two-terminal setup consisting of two copies (but with reversed chiralities) of the edge segment in {\bf (a)}.  
}
\label{supFig3:linejunctionwithleads}
\end{figure}

In this section, we perform a renormalization group (RG) analysis of the disordered aPf edge attached to leads, as depicted in Fig.~\ref{supFig3:linejunctionwithleads}\textcolor{blue}{(a)}. We perform our analysis near the two points $\theta = 0$ and $\theta = \pi$, see the discussion below Eq.~\eqref{eq:armso3rotatedbasis_new} in the main text. We will show that these points are RG fixed points: While the $\theta = 0$ fixed point (or equivalently, clean limit) is stable, $\theta = \pi$ is unstable. Our analysis is largely based on Ref.~\cite{Protopopov2017} where a $\nu=2/3$ edge was considered. Results of this section are used for the calculation of the QPC conductance in the coherent regime, as presented in the part ``Conductance in the coherent regime'' of the main text, with details of the calculation given in Sec.~\ref{sec:two-terminalconductance} of this Supplemental Material.

We consider an edge segment consisting of the second Landau level modes of the aPf edge structure, The segment is divided into three regions [see Fig.~\ref{supFig3:linejunctionwithleads}\textcolor{blue}{(a)}]: (i) the ``left lead'' $x < L/2$, (ii) the interacting and disordered region $|x| < L/2$, and (iii) the ``right lead'' $x > L/2$. The non-interacting ``bare modes'' $\phi_1$ and $\phi_{1/2}$ are the bosonic eigenmodes in regions (i) and (iii), while region (ii) is  described by the Lagrangian $\mathcal{L}_c + \mathcal{L}_n$ in Eq.~\eqref{eq:fixedpointLagrangian} in the main text. The charged boson $\phi_c$ is fully decoupled with the neutral boson $\phi_n $ and the Majorana mode $\chi$. Then, the total Hamiltonian for all regions (i), (ii), and (iii) reads  $H = H_{\theta = 0, \text{fix}} + H_{\text{pert}}$ with
\begin{align} \label{supeq:totalHam}
    H_{\theta= 0, \text{fix}} = & \int_{|x| >  L/2} \frac{dx}{4\pi} [ v_1(\partial_x \phi_1)^2 +2 v_{\frac{1}{2}}(\partial_x \phi_{\frac{1}{2}})^2] + 
    \int_{|x| <  L/2} \frac{dx}{4\pi} [ v_c(\partial_x \phi_c)^2 + \bar{v}_{n}(\partial_x \phi_{n})^2]  + 
     \int_{-\infty}^{\infty} dx \left ( \minus i \bar{v}_n \chi \partial_x \chi \right),
    \nonumber \\ 
     H_{\text{pert}} = &   \frac{2 i \theta \bar{v}_n}{\sqrt{\pi a}}  \sin (\phi_n (x=0)) \chi (x = 0) \,. 
\end{align}
As discussed in the main text, the transformation Eq.~\eqref{eq:transformationrepresentation2} in the main text collects all effects of disorder to the single point (which we choose as $x =0$), and the disorder strength is characterized by a single parameter $\theta$, which enters the term $H_{\text{pert}}$ in Eq.~\eqref{supeq:totalHam}.

We now identify the two values of $\theta$ where disorder can be effectively gauged away: $\theta= 0$ and $\theta = \pi$. For values $\theta \approx 0$, we can treat $H_{\text{pert}}$ as a weak perturbation and perform an RG analysis of $H_{\text{pert}}$ with respect to $H_{\theta= 0, \text{fix}}$. By contrast, for $\theta = \pi$, $H_{\text{pert}}$ can be fully absorbed into $H_{\theta=0,\text{fix}}$ by substituting $\phi_n \rightarrow \tilde{\phi}_n \equiv - \phi_n$ (thereby, $\phi_1 \rightarrow \tilde{\phi}_1$ and $\phi_{1/2} \rightarrow \tilde{\phi}_{1/2}$, see Eq.~\eqref{eq:tmatrix1pi} below for the exact mapping) and $\chi \rightarrow \tilde{\chi} \equiv - \chi$ in the $x> L/2$ region. With this transformation, the total Hamiltonian~\eqref{supeq:totalHam} turns into $H = H_{\theta = \pi, \text{fix}} + \tilde{H}_{\text{pert}}$ with
\begin{align} \label{supeq:totalHam2}
    H_{\theta= \pi, \text{fix}} = & \int_{x< \minus L/2} \frac{dx}{4\pi} [ v_1(\partial_x \phi_1)^2 +2 v_{\frac{1}{2}}(\partial_x \phi_{\frac{1}{2}})^2] + \int_{x > L/2} \frac{dx}{4\pi} [ v_1(\partial_x \tilde{\phi}_1)^2 +2 v_{\frac{1}{2}}(\partial_x \tilde{\phi}_{\frac{1}{2}})^2] \nonumber \\ &+ 
    \int_{|x| <  L/2} \frac{dx}{4\pi} [ v_c(\partial_x \phi_c)^2 + \bar{v}_{n}(\partial_x \phi_{n})^2] +
     \int_{-\infty}^{L/2} dx \left (\minus i \bar{v}_n \chi \partial_x \chi \right)
     +\int_{L/2}^{\infty} dx \left (\minus i \bar{v}_n \tilde{\chi} \partial_x \tilde{\chi} \right)
    \nonumber \\ 
     \tilde{H}_{\text{pert}} = &   \frac{2 i \tilde{\theta} \bar{v}_n}{\sqrt{\pi a}}
   \sin (\phi_n (x=0)) \chi (x = 0)\,, 
\end{align}
with $\tilde{\theta} \equiv \pi - \theta$. For $\theta\approx \pi \Leftrightarrow\tilde{\theta}\approx 0$, we can treat $\tilde{H}_{\text{pert}}$ as a weak perturbation to $H_{\theta= \pi, \text{fix}}$. 

Since the perturbation term $H_{\text{pert}}$ is local in $x$, we next integrate out the bosonic field $\phi_n (x)$ everywhere except at $x \neq 0$ to obtain an effective action only for $\phi_n (x = 0)$. This can be done within a transfer matrix approach which allows us to find the local Green's function for $\phi_n (x = 0)$. To do so, we first find the transfer matrix connecting the modes in different regions (i), (ii), (iii). For $\theta = 0$, the modes in the neighboring regions are connected by 
\begin{align} \label{eq:tmatrix1}
    \begin{pmatrix} 
    \phi_c^{\text{(ii)}} \\ \phi_n^{\text{(ii)}}
    \end{pmatrix}  = \mathcal{T} \begin{pmatrix} 
    \phi_1^{\text{(i)}}  \\ \sqrt{2} \phi_{1/2}^{\text{(i)}}
    \end{pmatrix}\,, \quad 
    \begin{pmatrix} 
    \phi_1^{\text{(iii)}}  \\ \sqrt{2} \phi_{1/2}^{\text{(iii)}}
    \end{pmatrix}  = \mathcal{T}^{-1} \begin{pmatrix} 
    \phi_c^{\text{(ii)}} \\ \phi_n^{\text{(ii)}}
    \end{pmatrix}  ,
    \quad \text{with}~\mathcal{T} = \frac{1}{T}\begin{pmatrix} 
     1 & R   \\ R & 1 
    \end{pmatrix}\,,
\end{align}
with $T = R = 1/\sqrt{2}$. These values come from the transformation between the ``bare mode'' basis and the ``charge-neutral'' basis. For $\theta = \pi$, the modes are instead connected by 
\begin{align}  \label{eq:tmatrix1pi}
    \begin{pmatrix} 
    \phi_c^{\text{(ii)}} \\ \phi_n^{\text{(ii)}}
    \end{pmatrix}  = \mathcal{T} \begin{pmatrix} 
    \phi_1^{\text{(i)}}  \\ \sqrt{2} \phi_{1/2}^{\text{(i)}}
    \end{pmatrix}\,, \quad 
    \begin{pmatrix} 
    \tilde{\phi}_1^{\text{(iii)}}  \\ \sqrt{2} \tilde{\phi}_{1/2}^{\text{(iii)}}
    \end{pmatrix}  = \mathcal{T}^{-1}_{\theta = \pi} \begin{pmatrix} 
    \phi_c^{\text{(ii)}} \\ \phi_n^{\text{(ii)}}
    \end{pmatrix}  ,
    \quad \text{with}~\mathcal{T}_{\theta = \pi} = \frac{1}{T}\begin{pmatrix} 
     1 & -R   \\ -R & 1 
    \end{pmatrix}\,. 
\end{align}
Note that for $\theta = \pi$, the transfer matrix connecting regions (ii) and (iii) requires special care; $\mathcal{T}_{\theta = \pi} \equiv \sigma_z \mathcal{T} \sigma_z$, where the Pauli-matrix $\sigma_z$ captures the additional minus sign that $\phi_n$ accumulates in the disordered region (ii). At the same time,  $\phi_c$ remains unaffected. Next, we combine $\mathcal{T}$ with $\mathcal{T}_{\theta=\pi}$ and define the following transfer matrix
\begin{align} \label{eq:tmatrix2}
\mathcal{T}_{\theta} =  \frac{1}{T}\begin{pmatrix} 
     1 & R e^{i \theta}   \\ R e^{-i \theta} & 1 
    \end{pmatrix}\,,
\end{align}
for $\theta = 0, \pi$. Then, the total transfer matrix
\begin{align} \label{eq:tmatrix3}
    \mathcal{T}_{\text{tot}} = \mathcal{T}_{\theta}^{-1} \mathcal{T}_{\text{dyn}} \mathcal{T}\,, \quad \text{with}\,\,\, \mathcal{T}_{\text{dyn}} = \begin{pmatrix} 
     e^{i \omega L / v_c} & 0   \\ 0 & e^{-i \omega L/v_n}
    \end{pmatrix}
\end{align}
bridges the modes in region (i) and the modes in the region (iii), and hence connects the incoming modes $\phi_{1/2}^\text{(i)}$ and $\phi_{1}^\text{(iii)}$, from $|x| \rightarrow \infty$ to the outgoing modes, $\phi_{1/2}^\text{(iii)}$ and $\phi_{1}^\text{(i)}$, propagating to $|x| \rightarrow \infty$. In Eq.~\eqref{eq:tmatrix3}, the matrix $\mathcal{T}_{\text{dyn}}$ takes into account the dynamical phase accumulated in region (ii). We now identify
\begin{align}
\label{eq:Tsqrt}
    \begin{pmatrix} 
    \phi_c (x = 0) \\ \phi_n (x= 0) 
    \end{pmatrix} = \sqrt{\mathcal{T}_{\text{dyn}}}\mathcal{T} \begin{pmatrix} 
    \phi_1^{\text{(i)}}  \\ \sqrt{2} \phi_{1/2}^{\text{(i)}}
    \end{pmatrix}\,,
\end{align}
where $\sqrt{\mathcal{T}{\text{dyn}}}$ comes from the fact that the distance between the (i)-(ii) interface, $x=-L/2$, and $x=0$ is $L/2$. From Eq.~\eqref{eq:Tsqrt},
we obtain $\phi_c (x = 0, \omega)$ and $\phi_n (x = 0, \omega)$, expressed in terms of the incoming modes as 
\begin{align}
    \phi_n (\omega, x = 0) = \frac{T e^{i \frac{\omega L}{2v_n}} (\sqrt{2}\phi_{1/2}^{\text{(i)}} + R e^{i L \omega / v_c} \phi_{1}^{\text{(iii)}} )}{1- e^{i \theta} R^2 e^{i \omega \tau}}\,, \quad 
    \phi_c (\omega, x = 0) = \frac{T e^{i \frac{\omega L}{2v_c}} (e^{i\theta} R \sqrt{2}\phi_{1/2}^{\text{(i)}} e^{i L \omega / v_n}  +  \phi_{1}^{\text{(iii)}} )}{1- e^{i \theta} R^2 e^{i \omega \tau}}.
\end{align}
Here, $\tau \equiv L / \bar{v}_{cn}$ is the average flight time in region (ii), with the mean velocity $\bar{v}_{cn} \equiv \frac{v_c v_n}{v_c + v_n}$. The local Green's function $g_{n}^{\theta} (\omega, x=0) \equiv  \langle \phi_n (\omega, 0) \phi_{n} (-\omega, 0)\rangle $ is then computed as 
\begin{align} \label{eq:Greenfunctionneutral}
    g_n^{\theta} (\omega, x=0) = \frac{T^2 (1 + R^2)}{1+ R^4 -  2 e^{i \theta} R^2 \cos (\omega \tau)} \frac{\pi}{\omega}\,,
\end{align}
where we used $\langle \phi_{1/2}^{\text{(i)}} (\omega, 0)  \phi_{1/2}^{\text{(i)}} (-\omega, 0) \rangle = \pi / (2\omega)$ and $\langle \phi_{1}^{\text{(iii)}} (\omega, 0)  \phi_{1}^{\text{(iii)}} (-\omega, 0) \rangle = \pi / \omega$, and $\phi_{1/2}^{\text{(i)}}$ and $\phi_{1}^{\text{(iii)}}$ are independent of each other.

With the Green's function~\eqref{eq:Greenfunctionneutral}, we are ready to perform the RG analysis close to $\theta = 0, \pi$. Near $\theta = 0$, 
we identify from the Hamiltonian~\eqref{supeq:totalHam} and the Green's function~\eqref{eq:Greenfunctionneutral} the effective action for $\phi_n (\omega, x = 0)$. It reads 
\begin{align} \label{eq:effectiveaction}
    S_n = \frac{1}{2}\int \frac{d \omega}{2\pi} \frac{ |\phi_n (\omega, x = 0)|^2}{|g_n^{\theta = 0} (\omega, x = 0)|} - \int d \tau \frac{2 i \theta \bar{v}_n}{\sqrt{\pi a}}
    \sin (\phi_n (x=0, \tau)) \chi (x = 0, \tau)\,.
\end{align}
By treating the second term of Eq.~\eqref{eq:effectiveaction} as a perturbation term, we obtain the tree level RG equation for $\theta$ as
\begin{align}
\label{eq:RG_Eq_1}
    \frac{d \theta}{d \ln \mathcal{L}} = \left (1 - \Delta_{\theta = 0} \right ) \theta \,, \quad \Delta_{\theta = 0} = \left (  \frac{T^2 (1 + R^2)}{2(1+ R^4 -  2  R^2 \cos (\omega \tau))} +  \frac{1}{2} \right)\,. 
\end{align}
Here, the running length scale $\mathcal{L} = a D/ \omega $ with the bandwidth $D$. Furthermore, $\Delta_{\theta = 0}$ is the scaling dimension of the perturbation term, where the term $1/2$ is the Majorana mode contribution. For sufficiently large $\omega > 1/ \tau$, the cosine term in $\Delta_{\theta = 0}$ oscillates rapidly in $\omega$ with period $2\pi / \tau$. The perturbation term is then effectively renormalized by the period-averaged value of $\Delta_{\theta = 0}$, i,e., $\frac{1}{2\pi}\int_{0}^{2\pi} d (\omega \tau) \Delta_{\theta = 0 } (\omega \tau) = 1$. With decreasing $\omega < 1/\tau$ (i.e., the dc-limit),  $\mathcal{L}$ starts to be affected by the leads and $\Delta_{\theta = 0}$ approaches $2$, upon inserting $R=T=1/\sqrt{2}$. 
In this frequency window, $\theta$ renormalizes to zero as $\mathcal{L}$ increases (or equivalently as $\omega$ decreases). Thus, $\theta = 0$ is a stable fixed point. 

For $\theta = \pi$, we find with an analogue analysis that the perturbation term in Eq.~\eqref{supeq:totalHam2} is governed by the RG equation 
\begin{align}
\label{eq:RG_Eq_2}
    \frac{d \tilde{\theta}}{d \ln \mathcal{L}} = \left (1 - \Delta_{\theta = \pi} \right ) \tilde{\theta} \,, \quad \Delta_{\theta = \pi} = \left (  \frac{T^2 (1 + R^2)}{ 2(1+ R^4 +  2  R^2 \cos (\omega \tau))} +  \frac{1}{2} \right)\,. 
\end{align}
Notice here the sign-difference in the denominator between $\Delta_{\theta=\pi}$ and $\Delta_{\theta =0}$ in Eq.~\eqref{eq:RG_Eq_1}. Now, when $\omega$ exceeds $1/\tau$, Eq.~\eqref{eq:RG_Eq_2} is effectively identical to Eq.~\eqref{eq:RG_Eq_1}. Crucially, however, with decreasing $\omega < 1/\tau$, $\Delta_{\theta = \pi}$ approaches $2/3$, which renders the perturbation RG-relevant. Hence, the perturbation grows upon normalization and the $\theta = \pi$ fixed point is thus unstable.

\section{Coherent regime: Two-terminal conductance in the absence of QPC and in the QPC geometry} 

\label{sec:two-terminalconductance}

In this section, we compute the two-terminal conductance in the coherent regime for two different setups: (A) The two-terminal conductance for the aPf state connected with leads (i.e., no QPC), shown in Fig.~\ref{supFig3:linejunctionwithleads}\textcolor{blue}{(b)}, and (B) the QPC setup depicted in Fig.~\ref{fig:APF_Setup} in the main text. 
This section contains details of the derivation presented in the part ``Conductance in the coherent regime'' of the main text.

The two-terminal setup in Fig.~\ref{supFig3:linejunctionwithleads}\textcolor{blue}{(b)} consists of three regions: (i) the ``left lead'' $x < L/2$, (ii) the interacting region $|x| < L/2$, and (iii) the ``right lead'' $x > L/2$. For this setup, we next compute the two-terminal conductance for all possible combinations of arm fixed points (i.e., $\theta_{i} = 0$ or $\pi$ with labels $i \in \{$U (up), D (down)$\}$). At the fixed points, the interacting region (ii) is described by 4 decoupled bosonic modes, $\phi_{c,\eta}$ and $\phi_{n, \eta'}$ propagating with chiralities $\eta = \pm $ and $\eta' = \pm$, respectively. The bosonic eigenmodes in the lead regions (i) and (iii) are $\phi_{1,\pm}$, $\phi_{1/2, \pm}$. The two neutral Majorana modes do not contribute to the conductance and hence we neglect them below. 

To compute the two-terminal conductance, we employ the transfer matrix approach as discussed above in Sec. The total transfer matrix bridging the two lead regions is given by 
\begin{align}
     \mathcal{T}_{\text{tot}} (\omega ) = \begin{pmatrix} 
     \mathcal{T}^{-1} & 0   \\ 0 & \mathcal{T}_{\theta_{\text{D}}}^{-1}
    \end{pmatrix} \begin{pmatrix} 
     \mathcal{T}_{\text{dyn}} & 0   \\ 0 & \sigma_x \mathcal{T}_{\text{dyn}}^{\dagger} \sigma_x
    \end{pmatrix} \begin{pmatrix} 
     \mathcal{T}_{\theta_{\text{U}}} & 0   \\ 0 & \mathcal{T} 
    \end{pmatrix}\,,
\end{align}
with $\mathcal{T}$, $\mathcal{T}_{\theta}, \mathcal{T}_{\text{dyn}}$ defined in Eqs.~\eqref{eq:tmatrix1}, \eqref{eq:tmatrix2}, and \eqref{eq:tmatrix3}. With the total transfer matrix, we find expressions for $\phi_c^+ (x = 0, \omega)$ and $\phi_c^- (x = 0, \omega)$ in terms of the incoming modes as
\begin{align} \label{eq:chargemodebcmatching}
    \phi_{c, +} (x = 0, \omega) &= \frac{T e^{i \frac{\omega L}{2v_c}} (e^{i\theta_{\text{U}}} R \sqrt{2}\phi_{1/2,-}^{\text{(iii)}} e^{i L \omega / v_n}  +  \phi_{1, -}^{\text{(i)}} )}{1- e^{i \theta_{\text{U}}} R^2 e^{i \omega \tau}}\,,  \nonumber \\ 
    \phi_{c, -} (x = 0, \omega) &= \frac{T e^{i \frac{\omega L}{2v_c}} (e^{i\theta_{\text{D}}} R \sqrt{2}\phi_{1/2,+}^{\text{(i)}} e^{i L \omega / v_n}  +  \phi_{1, -}^{\text{(iii)}} )}{1- e^{i \theta_{\text{D}}} R^2 e^{i \omega \tau}}\,,
\end{align}
with $R = T = 1/\sqrt{2}$. We next insert Eq.~\eqref{eq:chargemodebcmatching} into the conductance formula Eq.~\eqref{eq:kuboformulacond2}, derived below in Sec.~\ref{sec:kuboformula}, and use the Green's functions
\begin{align} \label{eq:relation}
      \langle [\phi_{1, \pm} (\omega_1), \phi_{1, \pm} (-\omega_1)] \rangle_{H_0}  = \frac{2\pi}{\omega_1}\,, \quad
      \langle [\phi_{1/2, \pm} (\omega_1), \phi_{1/2, \pm} (-\omega_1)] \rangle_{H_0} = \frac{\pi}{\omega_1}\,. 
\end{align}
Performing the frequency integration in Eq.~\eqref{eq:kuboformulacond2}, we arrive at the two-terminal conductance formula
\begin{align}
    G (\theta_{\text{U}}, \theta_{\text{D}}) = \frac{e^2}{2h} T^4 (1 + R^2) \left ( \frac{1}{(1-e^{i \theta_{\text{U}}} R^2)^2} + \frac{1}{(1-e^{i \theta_{\text{D}}} R^2)^2} \right),
\end{align}
with $\theta_{\text{U}}, \theta_{\text{D}} = 0, \pi$. Thus, the conductance (in units of $e^2/h$) for all possible combinations of arm fixed points takes the values
\begin{align} \label{eq:conductancelinejunction}
 G = \left \{ \begin{array}{lll}
 \frac{3}{2} & \,\, \text{for}\,\, \theta_{\text{U}} = \theta_{\text{D}} = 0 \,,
 \\ [2mm] \frac{5}{6} &   \,\,   \text{for} \ 
 \theta_{\text{U}} = 0 \,, \, \theta_{\text{D}} = \pi \ \text{or} \  \theta_{\text{U}} = \pi  \,, \, \theta_{\text{D}} = 0 \,, \\ [2mm]
 \frac{1}{6} &\,\, \text{for}\,\, \theta_{\text{U}} = \theta_{\text{D}} = \pi \,.
 \end{array} \right .
\end{align}

Next, we move on to the QPC setup, depicted in Fig.~\ref{fig:APF_Setup} in the main text. This setup consists of 5 regions: (i) the left lead $x < -L_{\text{QPC}}/2 - L$, (ii) the left arms of the QPC in $-L_{\text{QPC}}/2 - L < x < -L_{\text{QPC}}/2 $, (iii) the QPC region $|x| < L_{\text{QPC}}/2$, (iv) the right arms of the QPC in $L_{\text{QPC}}/2  < x <L + L_{\text{QPC}}/2 $, and (v) the right lead $x > L + L_{\text{QPC}}/2$. We now compute the conductance for all possible combinations of arm fixed points, i.e., for $\theta_{ij} = 0$ or $\theta_{ij}=\pi$ with labels $i\in\{$L\,(left), R\,(right)$\}$ and $j\in\{$U\,(up), D\,(down)$\}$, Fig.~\ref{fig:APF_Setup} in the main text. The total transfer matrix 
bridging regions (i) and (v) is given by
\begin{align}
    \mathcal{T}_{\text{tot}} (\omega ) = 
    \begin{pmatrix} 
     \mathcal{T}^{-1} & 0   \\ 0 & \mathcal{T}^{-1}_{\theta_{\text{RD}}}
    \end{pmatrix} 
    \begin{pmatrix} 
     \mathcal{T}_{\text{dyn}} & 0   \\ 0 & \sigma_x \mathcal{T}_{\text{dyn}}^{\dagger} \sigma_x
    \end{pmatrix}
    \begin{pmatrix} 
     \mathcal{T}_{\theta_{\text{RU}}} & 0   \\ 0 & \mathcal{T} 
    \end{pmatrix} 
     \mathcal{T}_{\text{QPC}}
    \begin{pmatrix} 
     \mathcal{T}^{-1} & 0   \\ 0 & \mathcal{T}_{\theta_{\text{LD}}}^{-1}
    \end{pmatrix} \begin{pmatrix} 
     \mathcal{T}_{\text{dyn}} & 0   \\ 0 & \sigma_x \mathcal{T}_{\text{dyn}}^{\dagger} \sigma_x
    \end{pmatrix} \begin{pmatrix} 
     \mathcal{T}_{\theta_{\text{LU}}} & 0   \\ 0 & \mathcal{T} 
    \end{pmatrix},
\end{align}
with $\mathcal{T}$, $\mathcal{T}_{\theta}, \mathcal{T}_{\text{dyn}}$ defined in Eqs.~\eqref{eq:tmatrix1}, \eqref{eq:tmatrix2}, and \eqref{eq:tmatrix3}. The additional matrix $\mathcal{T}_{\text{QPC}}$ is the transfer matrix in the QPC region, given by
\begin{align}
\label{eq:T_QPC}
   \mathcal{T}_{\text{QPC}} = \frac{1}{\sqrt{1- r^2}}  \begin{pmatrix} 
     1 & 0 & 0 & 0 \\ 0  & e^{i \theta_{\text{RU}}} & r e^{i (\theta_{\text{LD}} - \theta_{\text{RU}})} & 0 \\ 0  & - r & - e^{i \theta_{\text{LD}}} & 0 \\ 0 & 0 & 0 &1 
    \end{pmatrix}\,,
\end{align}
with $r$ a parameter to be taken to $1$ below. In Eq.~\eqref{eq:T_QPC}, the phase factors involving $\theta_{ij}$ reflect the phases accumulated in the arm regions. By using the same procedure as for Eqs.~\eqref{eq:chargemodebcmatching}-\eqref{eq:relation} and then letting $r\rightarrow 1$, we obtain Eq.~\eqref{eq:conductancecoh} in the main text. 

\section{Conductance range in the QPC geometry} \label{sec:conductancerange}

In this section, we use a  thermodynamic argument to show that the two terminal conductance $G$, for the QPC configuration depicted in Fig.~\ref{fig:APF_Setup} in the main text, is bound to range from $G = 2+ 1/17 = 35/ 17$ to $G = 2+ 1= 3$ (in units of $e^2 /h$).  These bounds agree with the results of calculation for QPC conductance at fixed points, see Eq.~\eqref{eq:conductancecoh} of the main text and Sec.~\ref{sec:two-terminalconductance} of this Supplemental Material.

\begin{figure}[t!]
\includegraphics[width =0.7\columnwidth]{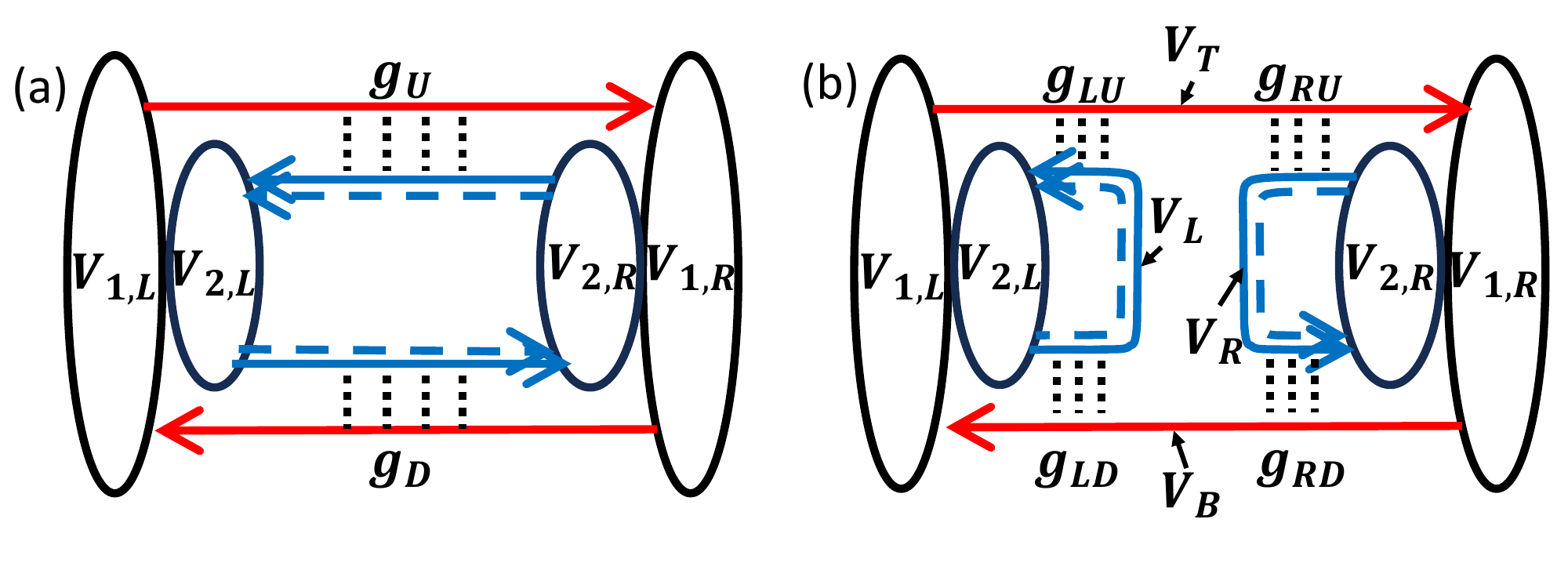}
\caption{{\bf Setups for the aPf state connected to leads}.  
{\bf (a)} The $\delta \nu = 1$ and $\delta \nu = -1/2$ bosonic edge modes are coupled to their respective contact. The upper edge modes are disconnected with the lower edge modes. {\bf (b)} The bosonic $\delta \nu = -1/2$ modes and the Majorana mode are fully reflected in the QPC region. This setup is identical with Fig.~\ref{fig:APF_Setup} in the main text when $V_{1, \text{L}} = V_{2,\text{L}}$ and $V_{1, \text{R}} =V_{2, \text{R}} $. The locations of voltages, referred to in Sec.~\ref{sec:conductancerange}, are marked out. 
}
\label{supFig4:condrange}
\end{figure}

We begin with considering a setup shown in Fig.~\ref{supFig4:condrange}\textcolor{blue}{(a)}. In contrast with the previous setups considered in this paper, the $\delta \nu = 1$ and $\delta \nu = -1/2$ modes are here each connected to fully independent reservoirs, characterized by the voltages $V_{1, \text{L}}$, $V_{1, \text{R}}$, $V_{2, \text{L}}$, and $V_{2, \text{R}}$. The upper and lower edge modes are completely disconnected to each other. 
The currents collected in the contact are then given by (we set here $e^2/h=1$)
\begin{align}
    \begin{pmatrix}
        I_{1, \text{L}} \\ I_{2, \text{L}} 
       \\ I_{2, \text{R}} \\ I_{1, \text{R}}
    \end{pmatrix} = \mathcal{G}
     \begin{pmatrix}
        V_{1, \text{L}} \\ V_{2, \text{L}} 
        \\ V_{2, \text{R}} \\ V_{1, \text{R}}
    \end{pmatrix}\,, \quad
    \text{with}\,\,
    \mathcal{G} = \begin{pmatrix} 
     -1 & g_{\text{D}} & 0 & 1- g_{\text{D}} \\ g_{\text{U}}  & -1/2 & 1/2 - g_{\text{U}} & 0 \\ 0  & 1/2 - g_{\text{D}} & - 1/2 & g_{\text{D}} \\ 1-g_{\text{U}} & 0 & g_{\text{U}} &-1 
    \end{pmatrix}\,.
\end{align}
Here, $g_{\text{U}}$ and $g_{\text{D}}$ are dimensionless coefficients that characterize back-scattering on the upper and lower edges, respectively. The total Joule heating power dissipated in all contacts is then given by $P = \sum_{i} I_i V_i = \sum_{i, j} \mathcal{G}_{ij} V_i V_j = \sum_{i, j} (\mathcal{G}_{ij} +\mathcal{G}_{ji})  V_i V_j /2$. The condition $P > 0 $ enforces that the symmetrized conductance matrix $\mathcal{G}_s \equiv (\mathcal{G} +\mathcal{G}^T)/2$ must be positive semi-definite, i.e., it has only non-negative eigenvalues. Explicitly, the eigenvalues of $\mathcal{G}_s$ read  $0, g_{\text{U}}+g_{\text{D}}, ( 3 - (g_{\text{U}}+g_{\text{D}}) \pm \sqrt{ 1 +  (g_{\text{U}}+g_{\text{D}})^2})/2 $. The condition of positive semi-definite $\mathcal{G}_{s}$ constrains $0 \leq g_{\text{U}}+g_{\text{D}} \leq 4/3$. Since $g_{\text{U}}$ and $g_{\text{D}}$ are fully independent, they can in principle be tuned independently, which leads to $0 \leq g_{\text{U}}, g_{\text{D}}\leq 2/3$. 

We note here that the constraint $0 \leq g_{\text{U}}, g_{\text{D}}\leq 2/3$, explicitly predicts that Andreev reflection-like behavior can occur in this system. To see this, we consider the single line junction setup depicted in Fig.~\ref{supFig3:linejunctionwithleads}\textcolor{blue}{(a)}. When the $\delta \nu = 1$ and $\delta \nu =  -1/2$ modes emanate from their respective contacts at the applied voltages, $V_1$ and $V_{2}$, the currents entering each contact, $I_1$ and $I_2$, are given by 
\begin{align} \label{eq:condmat}
    \begin{pmatrix}
        I_1 \\ I_2
    \end{pmatrix} = \mathcal{G} \begin{pmatrix}
        V_1 \\ V_2
    \end{pmatrix}\,, \,\, \text{with}\,\,\mathcal{G} =  \begin{pmatrix}
        1- g & g \\ g & \frac{1}{2} -g 
    \end{pmatrix}\,.
\end{align}
Here, $g$ is a dimensionless conductance parameter that follows the condition $0 \leq g \leq 2/3$ as above. From Secs.~\ref{sec:RGanalysis}-\ref{sec:two-terminalconductance} above, we note that $g = 0$ is the value corresponding to the clean fixed point ($\theta = 0$), while $g =2/3$ corresponds to the $\theta = \pi$ fixed point. 
This conductance matrix $\mathcal{G}$ in Eq.~\eqref{eq:condmat} shows that for $V_1 = 0$, $I_2 = (1/2 - g)V_2$. Therefore, $I_2$ can be negative, even for positive $V_2 >0$, provided that $1/2 < g \leq 2/3$. This negative conductance is reminiscent of an Andreev reflection process, since incoming electrons are reflected as holes while passing through the line junction. Such Andreev reflection-like behavior was proposed for the $\nu = 2/3$ edge~\cite{Protopopov2017}, and was further recently observed in a setup on an engineered interface with counter-propagating $\delta \nu = 1$ and $\delta \nu = -1/3$ modes~\cite{Hashisaka2021, Hashisaka2023}. 

With the condition $0 \leq g \leq 2/3$ and the conductance matrix Eq.~\eqref{eq:condmat}, we can now establish the permitted conductance range in the full QPC geometry, see Fig.~\ref{supFig4:condrange}\textcolor{blue}{(b)}. To do so, we apply Eq.~\eqref{eq:condmat} to each arm (LU, LD, RU, RD) of the QPC, 
\begin{align} \label{eq:voltagecurrent}
    \begin{pmatrix}
        V_{\text{T}} \\ I_{2,\text{L}}
    \end{pmatrix} = \mathcal{G}_{\text{LU}} \begin{pmatrix}
        V_{1,\text{L}} \\ V_\text{L}
    \end{pmatrix}\,, \quad  \begin{pmatrix}
        I_{1, \text{L}} \\ V_\text{L}/2
    \end{pmatrix} = \mathcal{G}_{\text{LD}} \begin{pmatrix}
        V_\text{B} \\ V_{2, \text{L}}
    \end{pmatrix}\,, \quad
      \begin{pmatrix}
        I_{1,\text{R}} \\ V_\text{R}/2
    \end{pmatrix} = \mathcal{G}_{\text{RU}} \begin{pmatrix}
        V_\text{T} \\ V_{2,\text{R}}
    \end{pmatrix}\,, \quad  \begin{pmatrix}
        V_\text{B} \\ I_{2, \text{R}}
    \end{pmatrix} = \mathcal{G}_{\text{RD}} \begin{pmatrix}
        V_{1, \text{R}} \\ V_\text{R}
    \end{pmatrix}\,. 
\end{align}
Here $I_{1, \text{L}}$, $I_{1, \text{R}}$, $I_{2, \text{L}}$, $I_{2, \text{R}}$ denote the currents entering contacts, $1\text{L}$, $1\text{R}$, $2\text{L}$, $2\text{R}$, respectively. By solving Eqs.~\eqref{eq:voltagecurrent} and identifying $V_{1,L} = V_{2, L} = V$ and $V_{1, R} = V_{2, R} = 0$ for the two-terminal setup, we obtain
\begin{align} \label{eq:twoterminalcond}
    G = \frac{I_{1, R} + I_{2, R}}{V} = 
    \frac{(-1 + 2 g_{\text{LD}} g_{\text{LU}}) (-1 + 2 g_{\text{RD}} g_{\text{RU}})}{1 - 4 g_{\text{LD}} g_{\text{LU}} g_{\text{RD}} g_{\text{RU}}}\,.
\end{align}
From the conditions $0 \leq g_{\text{LD}}, g_{\text{LU}}, g_{\text{RD}}, g_{\text{RU}} \leq  2/3$,we see that $G$ ranges from $1/17$ (obtained for $g_{\text{LD}}=g_{\text{LU}}=g_{\text{RD}}=g_{\text{RU}}=2/3$) to $1$. 
When the contribution of the two lowest-Landau-level integer modes is also taken into account, we thus see that $G$ ranges between $35/17$ and $3$. We finally note that $g_{ij} = 0$ corresponds to the value for the clean fixed point ($\theta_{ij} = 0$) while $g_{ij} =2/3$ corresponds to that for the $\theta_{ij} = \pi$ fixed point. By using these identifications for all $g_{ij}$, we immediately arrive at Eq.~\eqref{eq:conductancecoh} in the main text.

\section{Kubo formula} \label{sec:kuboformula}
In this section, we use linear response theory to derive a Kubo formula for the electric conductance of a FQH edge segment involving multiple edge modes.  The derived form of Kubo formula is used in Sec.~\ref{sec:two-terminalconductance} to calculate the QPC conductance in the coherent regime.

We consider $N$ ``bare'' edge modes described by the bosonic fields $\hat{\phi}_i$ with $1 \leq i \leq N$. The generic two-terminal conductance setup consists of three parts: (i) a left lead for $x < L/2$, (ii) the actual system in $|x| < L/2$, and (iii) the right lead $x > L/2$. The clean edge Hamiltonian
\begin{align}
    H_0 = \frac{1}{4\pi}\sum_{i,j= 1}^N\int dx\;  u_{ij}\partial_x \hat{\phi}_i (x) \partial_x \hat{\phi}_j (x),
\end{align}
includes both the kinetic terms with velocities $v_i\equiv u_{ii}$ and short-range density-density interactions $u_{ij}$ for $i\neq j$.  

We next analyze the electric current flowing in response to the time-dependent voltage $V (x, t)$. To properly model a two-terminal setup, we take this voltage to be constant in $x$ in the lead regions, i.e., $\partial_x V(x, t) = 0$ for $|x|> L/2$. More specifically, we take $V (x, t) = V_L (t)$ for $x < - L/2$ and $V (x, t) = V_R (t)$ for $x > L/2$. The voltage couples to the total, local charge densities
\begin{align} \label{eq:local density}
    \hat{\rho} (x) &= \frac{1}{2\pi} \sum_{i = 1}^N\partial_x \hat{\phi}_i (x) \,,
\end{align}
by adding to $\hat{H}_0$ the term
\begin{align} \label{eq:voltagecoupling}
    \hat{H}_V (t) &= -e \int dx \hat{\rho} (x) V (x, t) = - \sum_{i=1}^N \frac{e}{2\pi} \int dx \partial_x \hat{\phi}_i V(x, t) = -\frac{e}{2\pi} \int_{-L/2}^{L/2} dx ( - \partial_x V (x, t)) \sum_{i=1}^N \hat{\phi}_i (x) \,.
\end{align}
Within linear response, the average electric current $j (x, t)$ is given by  
\begin{align}
    j (x, t) = - i \int dt'\theta (t -t')\langle [\hat{j}(x, t), \hat{H}_V (t')] \rangle_{\hat{H}_0} = \frac{i e}{2\pi} \sum_{i=1}^N \int_{-L/2}^{L/2} dx' \theta (t -t') (-\partial_{x'} V(x', t')) \langle [\hat{j}(x, t), \hat{\phi}_i (x', t') ] \rangle_{H_0}\,,
\end{align}
from which we can identify the conductivity kernel $\sigma (x, x'; t-t')$, 
\begin{align} \label{eq:conductivity}
    \sigma (x, x'; t-t') = \frac{i e }{2 \pi} \theta (t - t') \sum_{i=1}^N  \langle [\hat{j} (x, t), \hat{\phi}_i (x', t') \rangle_{H_0}\,.  
\end{align}
Here, we used the definition of the conductivity kernel
\begin{align} \label{eq:linearresponse}
    j(x, t) \equiv \int_{-L/2}^{L/2} dx' \int dt' \sigma (x, x'; t-t') (-\partial_{x'} V(x',t'))\,.
\end{align}
In the frequency domain, the conductivity kernel~\eqref{eq:conductivity} becomes
\begin{align} \label{eq:conductivitykernel}
  \sigma (x, x'; \omega) = - \frac{e}{2\pi} \sum_{i=1}^N \int \frac{d\omega_1}{2\pi} \frac{1}{\omega_1 - \omega + i \eta}
  \langle [\hat{j} (x, \omega_1), \hat{\phi}_i (x', - \omega_1)] \rangle_{H_0}\,. 
\end{align}
Importantly, $\sigma (x, x')$ does not depend on $x$ and $x'$ in the dc limit ($\omega \rightarrow 0$) due to charge conservation. This independence allows us to move $\sigma$ outside the integration over $x'$ in Eq.~\eqref{eq:linearresponse}, which results in the formula
\begin{align}
    j (\omega) = \sigma (\omega) (V_L (\omega) - V_R (\omega))\,.
\end{align}
In the dc limit $\omega \rightarrow 0$, the dc conductance $G$ becomes
\begin{align} \label{eq:kuboformulacond}
    G &\equiv \lim_{\omega \rightarrow 0} \sigma (\omega) 
    = - \frac{e}{2\pi} \sum_{i = 1}^N \int \frac{d\omega_1}{2\pi} \frac{1}{\omega_1 + i \eta}
  \langle [\hat{j} (x, \omega_1), \hat{\phi}_i (x', -\omega_1)] \rangle_{H_0}\,. 
\end{align}
Since again $G$ does not depend on $x$ and $x'$, we can use this freedom to choose $x$ and $x'$ as we wish and thus choose $x = x' = 0$. Finally, we identify the current operator, $\hat{j}$, from the continuity equation $\partial_t \hat{\rho} + \partial_x \hat{j} = 0$ as     
\begin{align} \label{eq:currentoperator}
   \hat{j} (x, t) =  - \sum_{i=1}^N\frac{\partial_t \hat{\phi}_i}{2\pi}\,  \xrightarrow{FT} \hat{j} (x, \omega) = \sum_{i=1}^N \frac{i \omega}{ (2 \pi)} \hat{\phi}_i (x, \omega)\,,
\end{align}
which upon insertion into Eq.~\eqref{eq:kuboformulacond} relates $G$ to bosonic correlation functions via
\begin{align} \label{eq:kuboformulacond2}
    G = \frac{i e^2}{(2\pi)^2} \sum_{i, j = 1}^N \int \frac{d\omega_1}{2\pi} \frac{\omega_1}{\omega_1 + i \eta}
  \langle [\hat{\phi}_i  (x =0 , \omega_1), \hat{\phi}_j  (x = 0, -\omega_1)] \rangle_{H_0}\,. 
\end{align}
The remaining procedure to obtain $G$  is thus to compute the local correlation function $\langle [\hat{\phi}_i  (x =0 , \omega_1), \hat{\phi}_j  (x = 0, \omega_1)] \rangle_{H_0}$. This can be done by setting up transfer matrices that connect regions (i)-(iii). These matrices permit us to write $\hat{\phi}_i  (x =0 , \omega_1)$ in terms of the bosonic modes incoming from $|x| \rightarrow \infty$. In turn, this relation allows us to express the local correlation functions in terms of correlation functions of the incoming bosonic modes. We do this explicitly in Sec.~\ref{sec:two-terminalconductance} above. 
\end{document}